\documentclass[fleqn,usenatbib]{mnras}

\usepackage{newtxtext,newtxmath}
\usepackage[T1]{fontenc}

\DeclareRobustCommand{\VAN}[3]{#2}
\let\VANthebibliography\thebibliography
\def\thebibliography{\DeclareRobustCommand{\VAN}[3]{##3}\VANthebibliography}

\usepackage{graphicx}	% Including figure files
\usepackage{amsmath}	% Advanced maths commands
\usepackage{color,soul} % highlighting (temp)
\usepackage{verbatim}   % Comments (temp)

\usepackage{tikz,xcolor,hyperref}

\newcommand{\mgii}{Mg\,\textsc{ii}}
\newcommand{\civ}{C\,\textsc{iv}}
\newcommand{\feii}{Fe\,\textsc{ii}}

\newcommand{\nv}{N\,\textsc{v}}

\newcommand{\siiv}{Si\,\textsc{iv}}
\newcommand{\oiv}{O\,\textsc{iv}}

\newcommand{\oiii}{[O\,\textsc{iii}]}

\newcommand{\lya}{Ly\,\textsc{$\alpha$}}
\newcommand{\hbeta}{H\textsc{$\beta$}}

\newcommand{\mgiil}{Mg\,\textsc{ii}$\lambda$2799}
\newcommand{\civl}{C\,\textsc{iv}$\lambda$1549}

\newcommand{\nivl}{N\,\textsc{iv}]$\lambda$1486}

\newcommand{\siivl}{Si\,\textsc{iv}$\lambda$1398}
\newcommand{\oivl}{O\,\textsc{iv}$\lambda$1402}
\newcommand{\heiil}{He\,\textsc{ii}$\lambda$1640}
\newcommand{\oiiil}{[O\,\textsc{iii}]$\lambda$1663}

\newcommand{\hbetal}{H\textsc{$\beta$}$\lambda$4863}

\newcommand{\nvciv}{\nv/\civ}
\newcommand{\siivoivciv}{(\siiv+\oiv)/\civ}

\definecolor{lime}{HTML}{A6CE39}
\DeclareRobustCommand{\orcidicon}{%
    \begin{tikzpicture}
    \draw[lime, fill=lime] (0,0) 
    circle [radius=0.16] 
    node[white] {{\fontfamily{qag}\selectfont \tiny ID}};
    \draw[white, fill=white] (-0.0625,0.095) 
    circle [radius=0.007];
    \end{tikzpicture}
    \hspace{-2mm}
}

\newcommand{\orcidChrisO}{\href{https://orcid.org/0000-0003-0017-349X}{\orcidicon}}
\newcommand{\orcidChrisW}{\href{https://orcid.org/0000-0002-4569-016X}{\orcidicon}}
\newcommand{\orcidSamuel}{\href{https://orcid.org/0000-0001-9372-4611}{\orcidicon}}
\newcommand{\orcidFuyan}{\href{https://orcid.org/0000-0002-1620-0897}{\orcidicon}}
\newcommand{\orcidValentina}{\href{https://orcid.org/0000-0003-3693-3091}{\orcidicon}}
\newcommand{\orcidSebastian}{\href{https://orcid.org/0000-0003-0389-0902}{\orcidicon}}
\newcommand{\orcidGuido}{\href{https://orcid.org/0000-0002-6830-9093}{\orcidicon}}

\title[XQ-100 Black Hole Mass]{Virial Black Hole Mass Estimates of Quasars in the XQ-100 Legacy Survey}

\author[S. Lai et al.]{Samuel Lai,$^{1}$\orcidSamuel\thanks{E-mail: samuel.lai@anu.edu.au}
Christopher A. Onken,$^{1,2}$\orcidChrisO\,
Christian Wolf,$^{1,2}$\orcidChrisW\,
Fuyan Bian,$^{3}$\orcidFuyan\,
\newauthor{Guido Cupani,$^{4,5}$\orcidGuido\,
Sebastian Lopez$^{4}$\orcidSebastian\,
and Valentina D'Odorico,$^{5, 6, 7}$\orcidValentina\,
}
\\
$^{1}$Research School of Astronomy and Astrophysics, Australian National University, Canberra, ACT 2611, Australia\\
$^{2}$Centre for Gravitational Astrophysics, Research Schools of Physics, and Astronomy and Astrophysics, Australian National University, Canberra, ACT 2601, Australia\\
$^{3}$European Southern Observatory, Alonso de C\'{o}rdova 3107, Casilla 19001, Vitacura, Santiago 19, Chile\\
$^{4}$Departamento de Astronom\'{\i}a, Universidad de Chile, Casilla 36-D, Santiago, Chile\\
$^{5}$INAF – Osservatorio Astronomico di Trieste, Via G. B. Tiepolo 11, I-34143 Trieste, Italy\\
$^{6}$IFPU–Institute for Fundamental Physics of the Universe, via Beirut 2, I-34151 Trieste, Italy\\
$^{7}$Scuola Normale Superiore, Piazza dei Cavalieri, I-56126 Pisa, Italy\\
}

\date{Accepted XXX. Received YYY; in original form ZZZ}

\pubyear{2022}

\begin{document}
\label{firstpage}
\pagerange{\pageref{firstpage}--\pageref{lastpage}}
\maketitle

\begin{abstract}
The black hole (BH) mass and luminosity are key factors in determining how a quasar interacts with its environment. In this study, we utilise data from the European Southern Observatory Large Programme XQ-100, a high-quality sample of 100 X-shooter spectra of the most luminous quasars in the redshift range $3.5 < z < 4.5$, and measure the properties of three prominent optical and ultraviolet broad emission-lines present in the wide wavelength coverage of X-shooter: \civ, \mgii, and \hbeta. The line properties of all three broad lines are used for virial estimates of the BH mass and their resulting mass estimates for this sample are tightly correlated. The BH mass range is $\log{(\rm{M_{BH}}/\rm{M_\odot})} = 8.6-10.3$ with bolometric luminosities estimated from the 3000~\AA\ continuum in the range $\log{(\rm{L_{bol}}/\rm{erg\,s^{-1}})} = 46.7-48.0$. Robustly determined properties of these quasars enable a variety of follow-up research in quasar astrophysics, from chemical abundance and evolution in the broad-line region to radiatively driven quasar outflows. 
\end{abstract}


%%%%%%%%%%%%%%%%%%%%%%%%%%%%%%%%%%%%%%%%%%%%%%%%%%

%%%%%%%%%%%%%%%%% BODY OF PAPER %%%%%%%%%%%%%%%%%%

\section{Introduction}
Hundreds of thousands of quasar (QSO) sources have now been confirmed through massive surveys \citep[e.g.,][]{Flesch_2015_HMQ, Yao_2019, Lyke_2020} up to a redshift of $z = 7.642$ \citep{Wang_2021_z7.642}. Despite the abundance of sources, high-quality echelle spectroscopy is available for only a few thousand unique QSOs, of which only a fraction contain data in the near-infrared (NIR). As the redshift increases, more of the rest-frame ultraviolet (UV) and optical atomic transitions shift into the infrared, which renders NIR observations invaluable for QSO emission and absorption-line studies. 

The European Southern Observatory Large Programme ''Quasars and
their absorption lines: a legacy survey of the high-redshift universe with VLT/X-shooter" (hereafter referred to as XQ-100, PI: S. López, programme number 189.A-0424) is a publicly available and high-quality sample of echelle spectra from 100 luminous QSOs in the redshift range $3.5 < z < 4.5$ \citep{Lopez_2016_XQ100}. The simultaneous full spectral coverage is from 315 nm to 2500 nm with resolving power $R \sim 5400-8900$ and median signal-to-noise ratio (SNR) of 24, measured across the whole spectrum and entire sample of 100 QSOs. Prior to XQ-100, the largest NIR spectroscopic survey, conducted using the FIRE spectrograph at Magellan, was comprised of 50 QSOs at $2<z<5$ \citep{Matejek_2012} with a median SNR per-pixel of 13 across the entire QSO sample. The XQ-100 survey with its high SNR and broad spectral coverage provides a unique and statistically significant sample to study the rest-frame UV and optical spectral properties of 100 high-redshift QSOs. 

Among the scientific themes of the XQ-100 programme is the study of galactic absorption. Sub-damped (subDLA) or damped \lya\ systems \citep[DLA;][]{Wolfe_2005} are used to determine the cosmic density of neutral gas as they are the main reservoirs for neutral gas in the Universe \citep[e.g.,][]{Prochaska_2009, Noterdaeme_2012, Sanchez_2016, Berg_2019}. The same systems can be used to probe metal abundances of QSO hosts by tracing gaseous absorbers along QSO sightlines \citep{Berg_2016, Berg_2021}. Similarly, intrinsic narrow absorption lines (NALs) in XQR-30 data are probes of the physical conditions of the QSO immediate environment and energetics of its outflow \citep{Perrotta_2016}, where absorption-line diagnostics indicate metallicity, absorber covering fraction, and ionisation structure \citep{Perrotta_2018}. In addition, the XQ-100 spectra also addresses cosmological questions through independent constraints of the \lya\ forest power spectrum at high redshift \citep{Irsic_2017, Yeche_2017}.

The study of active galactic nuclei (AGN) properties is also one of the scientific themes from the XQ-100 programme. The high-quality spectra can be used for accurate measurements of $z > 3.5$ black hole masses using line profiles of rest-frame UV \civ, \mgii, or rest-frame optical \hbeta\ emission-lines and the continuum luminosity \citep[e.g.,][]{Mclure_2004, Greene_2005, Vestergaard_2006, Vestergaard_2009}. Flux ratios of emission-lines in the rest-frame UV, such as \nvciv\ or \siivoivciv, provide estimates of the metallicity in the QSO broad-line region (BLR), which probes the chemical enrichment history in high-redshift galactic nuclear regions \citep[e.g.,][]{Hamann_1999, Hamann_2002, Nagao_2006, Wang_2012, Xu_2018, Wang_2021, Lai_2022}. In the local universe, black hole masses and galactic bulge masses are strongly correlated \citep[the $\rm{M}_{\rm{BH}}-\rm{M}_{\rm{bulge}}$ relation;][]{Marconi_2003, Haring_2004, Greene_2010}, indicating that host galaxies and their central supermassive black holes co-evolve. Determining the black hole masses of high-redshift QSOs is valuable for studies that aim to investigate how properties of host galaxies and their black holes came to be strongly coupled \citep[e.g.,][]{Croton_2006, McConnell_2013, Terrazas_2020}.

In this study, we estimate the black hole masses of every source in XQ-100 using single-epoch virial estimates based on the prominent broad \civl\AA, \mgiil\AA, and \hbetal\AA\ lines. We measure emission-line properties utilising the high SNR, resolving power, and wide spectral coverage of the X-shooter data to tightly constrain the observed spectral profiles. This study produces a large catalogue of bright QSOs with robustly measured emission-line properties, black hole masses, and luminosity estimates at high-redshift ($z > 3.5$).

The content of this paper is organised as follows: in Section \ref{sec:sample}, we describe the XQ-100 data and their further processing. In Section \ref{sec:spectral_fitting}, we present our approach to modelling prominent emission-lines in the observed spectra. In Section \ref{sec:virial_mbh}, we describe virial mass estimates based on the measured line properties. In Section \ref{sec:results_discussion}, we discuss measurements of the emission-lines, black hole mass, and QSO luminosity. We compare the different virial mass estimates against each other and contextualise our results with large low-redshift samples. We summarize and conclude in Section \ref{sec:conclusion}. Throughout the paper, we adopt a flat $\Lambda$CDM cosmology with H$_{0} = 70$ km s$^{-1}$ Mpc$^{-1}$ and $\left(\Omega_{\rm m}, \Omega_{\Lambda}\right) = \left(0.3, 0.7\right)$. All referenced wavelengths of emission-lines are measured in vacuum.

\section{XQ-100 Sample Data and Processing} \label{sec:sample}
Targets in the XQ-100 sample were initially selected from the NASA/IPAC Extragalactic Database (NED) with declinations $\delta < +15^{\circ}$ and redshifts $z > 3.5$. An additional twelve targets were obtained from the literature with declination $+15^{\circ} < \delta < +30^{\circ}$. Deliberate steps were taken to avoid targets with known broad absorption features and to avoid intrinsic colour selection bias. A full description of the target selection process can be found in \citet{Lopez_2016_XQ100}. 

\subsection{Sample Description and Data Reduction}
The targets span the redshifts from $z = 3.508$ to $z = 4.716$ \citep{Lopez_2016_XQ100}, although all but four are within the redshift range $3.5 < z < 4.5$. The sample is biased towards bright sources, covering a magnitude range in \textit{Gaia DR3} G$_{\rm{RP}}$ band \citep{GaiaEDR3_2021} of 16.78 to 19.00 Vegamag. Observations were carried out between 2012 April 1, and 2014 March 26 by the X-shooter instrument \citep{Vernet_2011_Xshooter} on the Very Large Telescope (VLT) using all three spectroscopic arms: UVB (300$-$559.5 nm), VIS (559.5$-$1024 nm), and NIR (1024$-$2480 nm). The wide wavelength coverage ensures that the \civ\ and \mgii\ emission-lines are always observed within the VIS and NIR arms for the range of redshifts in the sample. Additional information on the requested observing conditions and instrumental setup is available in \citet{Lopez_2016_XQ100}. We briefly summarise the reduction and processing procedures behind the XQ-100 data products, as described in \citet{Lopez_2016_XQ100}.

Extraction of XQ-100 spectra was performed using an IDL-based custom pipeline \citep{Becker_2012}. The strategy of the custom pipeline follows techniques described in \citet{Kelson_2003}. Flux calibration uses response curves generated from observations of spectro-photometric standard stars, observed close in time to the science frames \citep{Lopez_2016_XQ100}, where a fiducial response curve was used if the temporally closest standard star observation was not optimal. Newer versions of this pipeline have been used in other QSO studies, such as XQR-30 \citep{Dodorico_2023_XQR30}. While XQ-100 data from all three spectrograph arms are available, for the present study, we consider only the VIS and NIR arms, because they contain all emission-lines of interest. The velocity resolution chosen to rebin the spectra are 11 km s$^{-1}$ and 19 km s$^{-1}$ for the VIS and NIR arms, respectively.

The absolute flux calibration is a crucial step in determining the luminosity of the QSO continuum. A comparison between XQ-100 and Sloan Digital Sky Survey \citep[SDSS;][]{York_2000_SDSS} spectra showed a systematic underestimation of flux for the X-shooter spectra due to slit losses. However, the slit losses appear to be roughly achromatic, such that the spectral shape is correctly reconstructed, but the flux calibration should be taken as order-of-magnitude estimates \citep{Lopez_2016_XQ100}. Thus, we describe our independent calibration of the XQ-100 spectra to observed photometry in Section \ref{sec:data-processing}.

Telluric absorption features appear prominently in both the VIS and NIR arms. Corrections to the spectra are derived using model transmission spectra based on the ESO SKYCALC Cerro Paranal Advanced Sky Model, version 1.3.5 \citep{Noll_2012, Jones_2013}, which are applied to individual-epoch spectra of all XQ-100 QSOs. After extraction and telluric correction, the median per-pixel SNR for the whole QSO sample are 33, 25, and 43, measured at rest-frame wavelengths 1700, 3000, and 3600 \AA, respectively \citep{Lopez_2016_XQ100}, computed in $\pm 10$\AA\ windows.

The processed XQ-100 data products, including reduced spectra and telluric models, are publicly available through the \href{http://archive.eso.org/wdb/wdb/adp/phase3_main/form?collection_name=XQ-100&release_name=DR1}{ESO Science Archive Facility}. However, the spliced spectra and the multi-epoch averaged spectra are not telluric corrected.

\subsection{Data Post-Processing} \label{sec:data-processing}
We obtain individual VIS and NIR single-epoch frames from the ESO Science Archive Facility for all XQ-100 sources and apply the following post-processing procedure:
\begin{enumerate}
    \item We use the respective telluric model included in each frame to obtain the telluric-corrected spectra and use the emission redshift to transform the spectra into the rest-frame.
    \item We identify pixels for which the per-pixel SNR is 5 or below and mask them from further processing and modelling.
    \item We apply a mask by sigma-clipping with a 3$\sigma$-threshold along a box width of 40 pixels to remove some of the narrow absorption features and noise above 3$\sigma$. The absorption features are not desired when modelling the intrinsic profile of the broad emission-lines and the sigma-clipped spectrum also helps constrain the continuum. While this procedure alone will not remove the base of absorption troughs, we follow the procedure in \citet{Shen_2011}, which defines our single-epoch virial mass calibration of \mgii. In Section \ref{sec:line-fitting}, we describe an additional mask buffer window to remove the base of absorption features embedded in the \civ\ line profile, but this is not applied throughout the entire spectrum.
    \item We crossmatch the XQ-100 sample with UKIRT Infrared Deep Sky Survey \citep[UKIDSS;][]{UKIDSS} DR11, UKIRT Hemisphere Survey \citep[UHS;][]{UHS_DR1} DR1, VISTA Hemisphere Survey \citep[VHS;][]{VHS} DR6, VISTA Kilo-degree Infrared Galaxy Survey \citep[VIKING;][]{VIKING} DR5, and Two Micron All-Sky Survey \citep[2MASS;][]{2MASS} to obtain near-infrared \textit{J}-band photometry. We also crossmatch all targets with the SkyMapper Southern Survey \citep[SMSS;][]{SMSS_2019} DR3, Panoramic Survey Telescope and Rapid Response System \citep[Pan-STARRS;][]{PanSTARRS} DR1, Sloan Digital Sky Survey \citep[SDSS;][]{York_2000_SDSS} DR16, and Dark Energy Sky Survey \citep[DES;][]{DES_2021} DR2 to obtain optical \textit{i}-band photometry. We obtain the transmission profile of the broadband filters using the SVO Filter Profile Service \citep{SVO_Filter_Profile_Service} and integrate the observed-frame spectra across the profile, obtaining a flux ratio between the photometry and spectrum with an associated uncertainty, which is used to calibrate the observed spectra to the photometry. There is one target, SDSS J004219.74$-$102009.4, for which no publicly available \textit{J}-band photometry was found in the above surveys. In this case, we scale the flux of the NIR arm to match the flux of the VIS arm within the overlapping wavelength coverage. The magnitudes used for calibration are provided in the online supplementary table. We note that the median correction required to match the spectrum to photometry is a 42\% flux increase with an error of 2--3\%, which is higher than the $\sim$ 30\% flux underestimation on X-shooter's part compared to SDSS spectra estimated in \citet{Lopez_2016_XQ100}. As the photometry is taken from a separate epoch from the spectroscopic data, the additional uncertainty from the photometric calibration is insignificant compared to QSO variability, which we quantify and discuss in Section \ref{sec:variability}.
    \item We standardise the rest-frame wavelength domain for all of the spectra. Every spectrum is resampled using a flux-conserving algorithm \citep[\texttt{SPECTRES};][]{Carnall_2017} into rest-frame bins with a common velocity dispersion of 50 km s$^{-1}$. The resampling calculation and error propagation are described in detail in \citet{Carnall_2017}. Then the VIS and NIR arms are spliced together without rescaling, using the inverse variance weighted mean flux for the superposition between arms. In a few cases, we observe a discontinuity between the VIS and NIR arms, located between 1860\AA\ to 2275\AA\ for the redshift range of our sample. The median flux difference between arms as measured in the overlapping region is 0.6\%, albeit with a large standard deviation of 24\%. However, we emphasise that the data in the overlapping region between arms are naturally at the edge of the wavelength coverage of each arm and is particularly noisy, so the flux difference measured in this fashion can be exaggerated. Nevertheless, we flag all targets with higher than 25\% flux difference between the VIS and NIR arms in the supplementary table under the column ``NIR\_VIS\_Flag''. We rely on the flux calibration in each respective arm and only use data within one arm at a time to fit the QSO continuum. Thus, the flux discontinuity between arms does not affect our continuum or emission-line models.
    \item If there are repeated observations of a single source, we make use of all the available data and stack the resampled telluric-corrected spectra together, using the mean weighted by the inverse variance to define the value at each 50 km s$^{-1}$ velocity bin and propagate the uncertainty. Because of the calibration in step (iv), the flux density at each velocity bin between repeated observations are in good agreement. The temporal separation between repeated observations range from 10 days to 1.5 years. However, we are interested in the average spectrum in order to determine representative properties of the black hole mass and luminosity. Prior to rescaling the flux level of the spectra to photometry, the median flux difference between exposures measured at every velocity bin is 7.3\% with a standard deviation of 7.6\%. After rescaling, our flux level is more consistent, measured at 2.3\% with a standard deviation of 0.6\%. We also quantify the uncertainty from QSO variability in Section \ref{sec:variability}.
    \item We use $R_{\rm{v}} = 3.1$ and the Schlegel, Finkbeiner \& Davis \citep[SFD;][]{Schlegel_1998} extinction map to apply a correction for the Milky Way extinction in the observed frame. However, the normalisation of the colour excess based on the SDSS footprint and fits to the blue tip of the stellar locus suggests that SFD systematically over-predicts $E(B-V)$ by 14\%. \citep{Schlafly_2010} Thus, we apply a 14\% re-calibration factor to the colour excess, such that $E(B-V) = 0.86 \times E(B-V)_{\rm{SFD}}$ \citep{Schlafly_2011}.
    
\end{enumerate}

After the post-processing procedure, the median SNR per 50 km s$^{-1}$ for the whole QSO sample measured at rest-frame 1700, 3000, and 3600 \AA\ is 76, 52, and 74, respectively, measured from the median SNR within $\pm10$\AA\ windows. Much of the increase in signal originates from the SNR floor and consolidating the flux from its native resolution into the rest frame 50 km s$^{-1}$ grid. 

For the wavelength range redder than rest-frame 3600 \AA, a similar post-processing procedure is applied, but the sigma-clip mask of step (iii) is omitted to preserve narrow emission-line features of \hbeta\ and \oiii. Due to the redshift range of this sample, only a subset of sources contains the \hbeta\ line within the X-shooter coverage. We visually inspect the data to ensure that the \hbeta\ line is distinguishable from the additional noise of the thermal background and second-order contamination at edge of the NIR arm wavelength coverage. We also ensure that the \hbeta\ line is observed with sufficient SNR (> 10 per resolution element), which produces a sub-sample of 21 QSOs, where the median 50 km s$^{-1}$ SNR across all 21 QSOs is 13. In this case, the SNR of each QSO is measured from the median SNR between 5090$-$5110 \AA.  

\section{Spectral Modelling} \label{sec:spectral_fitting}
Our objective in this study is to measure the properties of the following QSO broad emission-lines: \civl\AA, \mgiil\AA, and \hbetal\AA. In the XQ-100 sample, both \civ\ and \mgii\ can be located in all spectra, while \hbeta\ is observable only in lower redshift targets with sufficient signal. In this section, we describe our approach towards modelling emission-lines, using a publicly available code \citep[\texttt{PyQSpecFit}\footnote{\hyperlink{https://github.com/samlaihei/PyQSpecFit}{https://github.com/samlaihei/PyQSpecFit}};][]{PyQSpecFit_v1} designed specifically for modelling QSO spectral lines.

\subsection{Continuum Modelling} \label{sec:cont-fitting}

Although a continuum model is provided as part of the XQ-100 data products, we elect to use our own continuum model due to how sensitive the broad emission-line models are to the local continuum. Our model follows similar studies \citep[e.g.,][]{Wang_2009} in that the underlying continuum is built from two components: a power-law continuum and \feii\ template, simultaneously fit to selected pseudo-continuum-modelled wavelength regions. We briefly comment on the Balmer continuum later in this section and quantify its effect in Appendix \ref{sec:appendix-Balmer}. All components of the pseudo-continuum are used in measuring the \mgii\ and \hbeta\ emission-lines, but the flux contribution from the \feii\ continuum is less significant in the wavelength region of \civ. Thus, we only use a power-law to constrain the continuum in the vicinity of the \civ\ line.

The power-law continuum is defined by the following function normalised at rest-frame 3000 \AA, 
\begin{equation}
    F_{\rm{pl}}(\lambda; F_{\rm{0}}, \gamma) = F_{\rm{pl, 0}} \left(\frac{\lambda}{3000 \mbox{\normalfont\AA}}\right)^{\gamma}\,,
    \label{eq:pl-cont}
\end{equation}
where $F_{\rm{pl, 0}}$ and $\gamma$ are the normalization and power-law slope, respectively. 

The \feii\ continuum is of considerable importance to the \mgii\ and \hbeta\ models, as both lines are sensitive to the features of the \feii\ contribution underneath the emission-line. To eliminate the \feii\ emission when strong, we convolve the \feii\ model with a Gaussian broadening kernel $G(\lambda, \sigma)$ of standard deviation $\sigma$ in order to match the variety of features observed in our spectra. The Gaussian broadening follows,
\begin{equation}
    F_{\rm{Fe}}(\lambda; \zeta_{\rm{0}}, \delta, \sigma) = \zeta_{\rm{0}} \,  F_{\rm{template}}|_{\lambda(1+\delta)} \circledast G(\lambda, \sigma)\,,
\end{equation}
where the free parameters of the \feii\ contribution include the flux scaling factor denoted by $\zeta_{\rm{0}}$, the standard deviation of the broadening kernel $\sigma$, and a small multiplicative wavelength shift $\delta$. Furthermore, we consider a variety of empirical and semi-empirical \feii\ emission templates: \citet[][VW01]{Vestergaard_2001} and \citet[][M16]{Mejia-Restrepo_2016} cover the rest-frame UV while \citet[][BG92]{Boroson_Greene_1992} and \citet[][P22]{Park_2022} cover the rest-frame optical. \citet[][BV08]{Bruhweiler_Verner_2008} and \citet[][T06]{Tsuzuki_2006} cover both regions. We use a VW01 template spliced with the \citet{Salviander_2007} template, which extrapolates underneath the \mgii\ line from rest-frame 2200$-$3090\AA. Furthermore, the wavelength range 3090$-$3500\AA\ is augmented with the T06 template \citep{Shen_2012}. This version of VW01 is also used in other QSO modelling codes such as \texttt{PyQSOFit} \citep{Guo_2018}. We find the typical value of the Gaussian broadening dispersion $\sigma$ to be 1600 km s$^{-1}$ in the rest-frame UV and 1300 km s$^{-1}$ in the rest-frame optical. 

\begin{figure}
	\includegraphics[width=1.0\columnwidth]{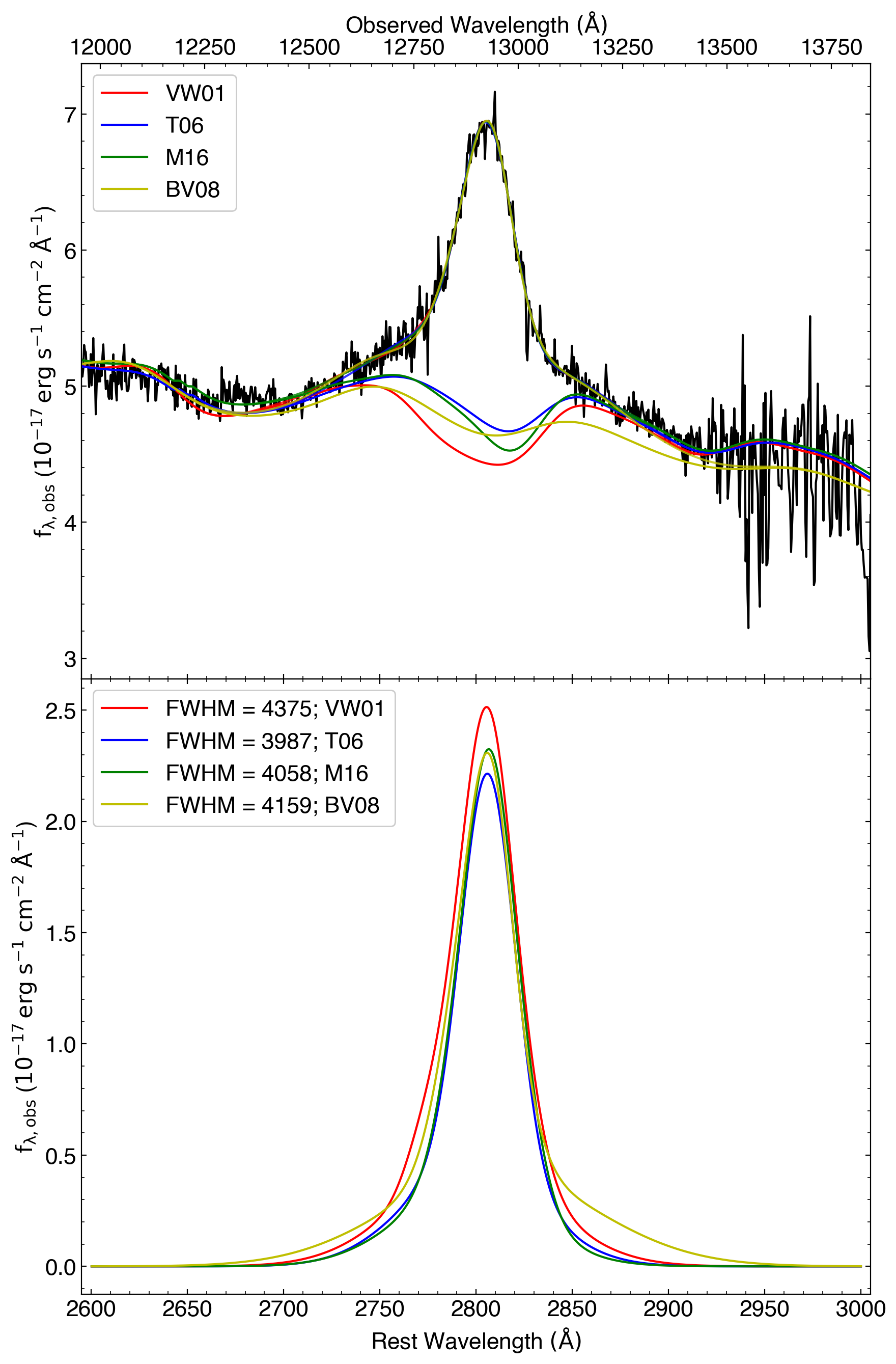}
    \caption{Example model of the \mgii\ emission feature of J110352$+$100403 with the combined pseudo-continuum from the power-law, Balmer, and \feii\ components. The line models differ from one another by the applied \feii\ template as indicated in the legend. The resulting continuum for each template is plotted in solid lines in the top panel and the continuum-subtracted line profile model is plotted in the bottom panel. In this example, the \mgii\ FWHM between models with different \feii\ templates ranges from 4000 to 4400 km s$^{-1}$.}
    \label{fig:Fe_templates_sample}
\end{figure}

In this work, we are not concerned with the specific properties of the \feii\ emission, and thus we will not discuss the physical interpretation of the dispersion and velocity shifts of the \feii\ emission. The \feii\ pseudo-continuum is used solely as an approximation to remove iron emission when significant in the spectra. Figure \ref{fig:Fe_templates_sample} shows the spectrum of J110352$+$100403 close to the \mgii\ region with the four UV \feii\ templates models overplotted along with the resulting emission-line models in a separate panel. Properties of the \mgii\ line model depend sensitively on the assumed \feii\ model. In the extreme case of SDSS J093556.91$+$002255.6, differences in the \feii\ model alone are responsible for shifting the measured full-width half-maximum (FWHM) in a range from 3700 to 5200 km s$^{-1}$. Similarly, the \hbeta\ line model is also sensitive to the optical \feii\ model. In Section \ref{sec:line-fitting}, we discuss how the differences in measured line properties resulting from various \feii\ models inform the measurement uncertainty.

The full pseudo-continuum is the sum of all contributing components, which is uniquely defined by 5 free parameters. All components of the continuum are fit simultaneously to selected pseudo-continuum windows close to the emission-line of interest. For each emission feature, the local underlying continuum is fit separately. We do not fit a ``global" continuum across the the spectral range from \civ\ to \hbeta\ in order to avoid biases due to deviations from a single power-law model, such as dust reddening \citep[e.g.,][]{Richards_2003} and host galaxy contributions \citep[e.g.,][]{VandenBerk_2001}. Outside irregular circumstances, such as a discontinuity in the flux-calibrated spectrum between the VIS and NIR arms, the pseudo-continuum modelling windows are selected from: 1275--1290\AA, 1348--1353\AA, 1445--1455\AA, 1687--1697\AA, 1973--1983\AA, 2200--2750\AA, 2820--3300\AA, 3500--3800\AA, 4200--4230\AA, and 4435--4700\AA\ with occasional $\pm30$\AA\ deviations to suit specific features of the spectra, avoid telluric regions, or to accommodate the properties of particular emission or absorption features.

The Balmer continuum is also often included in the pseudo-continuum model when modelling QSO spectra, but it is not always well-constrained and is degenerate with the power-law and \feii\ continuum \citep[e.g.,][]{Wang_2009, Shen_2012}, such that the Balmer contribution is not considered for the underlying continuum in some other QSO studies \citep[e.g.,][]{Shen_2011}. For the XQ-100 sample, we find that the Balmer continuum properties are not well-constrained and the broad emission-line decomposition is not strongly affected by either the inclusion or exclusion of the Balmer continuum. Therefore, in the following sections, we present our results without the Balmer continuum, but we quantify the effect of its inclusion in Appendix \ref{sec:appendix-Balmer}.

\subsection {Line Modelling} \label{sec:line-fitting}
Broad emission-line profiles exhibit a wide range of properties and complexities from asymmetries to multiple peaks and plateaus, making single Gaussian models unsuitable. Instead, many QSO spectral modelling studies use a multiple Gaussian approach to fit each emission feature \citep[e.g.,][]{Greene_2005, Shen_2011, Rakshit_2020}. Following these studies, we fit each broad emission-line with multiple $(N_{\rm{gauss}})$ symmetric Gaussian functions having $3 \times N_{\rm{gauss}}$ free parameters, to obtain smooth realisations of the observed line profile. Similar to \citet{Rakshit_2020}, we choose $N_{\rm{gauss}}$ = (3, 4, 4) for \civ, \mgii, and \hbeta, respectively. The 4 Gaussian components of the \mgii\ and \hbeta\ lines are divided into 3 broad and 1 narrow component. For high luminosity QSOs, such as the targets in the XQ-100 sample, \oiii\ lines with FWHMs exceeding 1000 km s$^{-1}$ are more common \citep[e.g.][]{Shen_2012, Coatman_2019}, so we adopt a FWHM upper threshold of 1200 km s$^{-1}$ for the \hbeta\ narrow lines. The \mgii\ narrow lines are often ambiguous and poorly constrained. Thus, we further constrain the upper FWHM threshold for \mgii\ to 1000 km s$^{-1}$, ensuring that the modelled components are indeed narrow. To measure the broad-line properties of each line, we subtract the narrow-line contribution from the total line profile, using only the 3 broad components to model the emission-line.

\begin{figure*}
\begin{tabular}{c}
  \includegraphics[width=\textwidth]{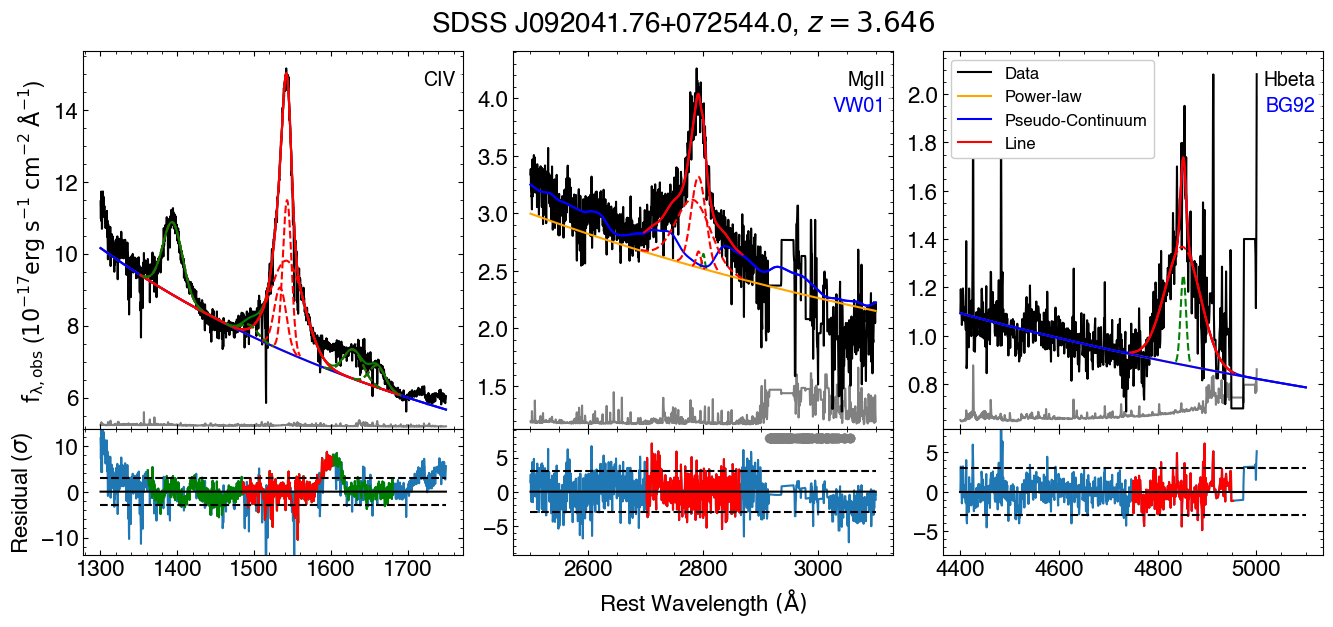} \\   \includegraphics[width=\textwidth]{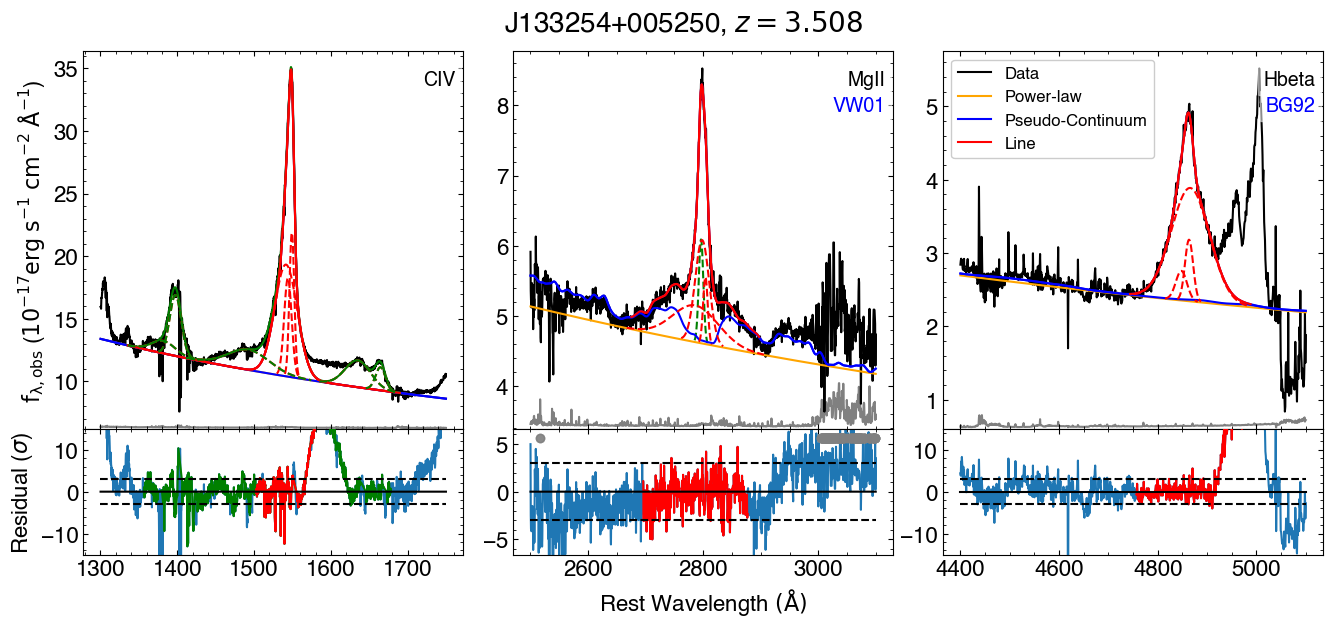} 
\end{tabular}
\caption{Example models of the \civ, \mgii, and \hbeta\ emission-lines from SDSS J092041.76$+$072544.0 and J133254$+$005250. The spectroscopic data are plotted in black, its error spectrum in grey, and the power-law continuum is in orange. The error spectrum is shifted vertically such that the bottom of the panel represents zero flux error. The combined pseudo-continuum which includes a power-law and the \citet{Tsuzuki_2006} \feii\ template is plotted in blue. The red lines indicate the total line profile and the dashed lines are the multiple Gaussian decomposition. We show the narrow-line model with the green dotted line and in the \civ\ panel, green highlights the line models of neighbouring lines. The residuals are shown normalized by the flux error, $\sigma$, in order to represent the quantity minimised by the line-modelling algorithm, (data-model)/error. We represent $\pm3\sigma$ in the residual panel with the dashed lines and we also show the result of the automated sigma-clipping in the \civ\ residual, where masked features remain in blue rather than red or green. Telluric absorption windows are denoted by the solid grey bar in the residual panel. These figures are available for all targets in the online supplementary material.} \label{fig:Example_fits}
\end{figure*}

The adjoining \oiii\ lines can present a challenge for modelling the redder wing of the \hbeta\ line profile, but they are also useful to constrain the width of the narrow \hbeta\ component. However, these adjoining lines are infrequently detected in the XQ-100 spectra with J133254$+$005250 presented in Figure \ref{fig:Example_fits} as one of the only two cases with detectable \oiii, alongside J101818$+$054822. Without the presence of the \oiii\ lines, the decomposition of the total \hbeta\ line profile into its broad and narrow emission may not be unique.

While all broad emission-line profiles are affected by embedded narrow absorption features, the effect that they have on the resulting model is more significant for \civ. In order to obtain more appropriate models of the intrinsic \civ\ broad emission profile, we apply an additional 2500 km s$^{-1}$ box-width sigma-clip mask with a 3$\sigma$-threshold. Every contiguous masked region has a masked buffer window of 3 pixels, equivalent to 150 km s$^{-1}$, applied on each end. Most broad \civ\ emission profiles can be fit automatically, but some QSOs contain features which are visually inspected and masked. 

If not accounted for, neighbouring lines can influence the measurement of the intrinsic \civ\ broad-line properties. In order to disentangle the \civ\ line model from its neighbours, we simultaneously model the broad \siivl\AA, \oivl\AA, \nivl\AA, \heiil\AA, and \oiiil\AA\ lines, using one broad component or one broad and one narrow component to constrain each neighbouring line. We find that the \civ\ line properties are not sensitive to whether narrow features, if present, in adjacent lines are modelled as a separate component. The neighbouring lines are only modelled in order to account for their influence on the \civ\ line properties. As such, we do not tabulate properties of the neighbouring lines in our catalogue.

For both \mgii\ and \hbeta, the final emission-line properties, tabulated in Table \ref{tab:XQ100_properties_sample}, are determined as the average of properties measured from the resulting line models, created by applying in turn each of the four \feii\ templates. We consider two primary sources of uncertainty in the measurement of emission-line and continuum properties. One source is the uncertainty from the various \feii\ emission templates and another is the measurement uncertainty. We estimate the uncertainty from the \feii\ template by independently modelling each spectrum with four models; VW01, T06, BV08, and M16 at UV wavelengths and BG92, T06, P22, and BV08 at optical wavelengths. Then we measure the line properties for a given model and quote the standard deviation. We also estimate the measurement uncertainty using a Monte Carlo approach by creating 50 synthetic spectra for individual target \citep[e.g.,][]{Shen_2011}, where the flux at each pixel is resampled from a symmetric distribution with a standard deviation equivalent to the pixel flux error. We assume that the noise in the spectrum follows a normal distribution. After modelling all of the synthetic spectra independently and varying the \feii\ template, the final measurement uncertainty is determined from these two sources added in quadrature. Each of the two sources contributes roughly equivalent uncertainty to emission-line FWHM, but the choice of \feii\ emission template dominates the variance in the measured continuum luminosity. 

For the \civ\ line properties, we use only the Monte Carlo uncertainty estimated by modelling 50 synthetic spectra. In this case, we do not consider the uncertainty from modelling different \feii\ templates, because the \feii\ emission is weak in this wavelength region and it is not used to define the continuum. The different realisations of the resampled spectra and their resulting line models help to capture degeneracies in the way that flux can be distributed between \civ\ and its neighbouring lines, propagating that degeneracy into the uncertainty of the line properties.

We provide a data quality flag, indicated by ``Quality\_Flag'', for each emission-line model which is used to identify where the median SNR per 50 km s$^{-1}$ resolution element of the data within the emission-line modeling region is below 20. We manually flag additional targets with the \mgii\ data quality flag to indicate poor quality fits or that significant residual telluric features are evident in the spectrum. As there are no targets below the SNR threshold for the data quality flag for \civ, we instead use the flag to indicate where we have manually adjusted the fit, by choosing more appropriate continuum windows or manually masking absorption features. Additionally, we flag targets, using ``Hbeta\_Truncation\_Flag'', for which the red wings of the \hbeta\ profile is clearly truncated by the edge of X-shooter's NIR arm wavelength coverage, which would reduce the reliability of the line model. However, we do not exclude flagged targets from further analysis and contextualisation of the XQ-100 sample in Section \ref{sec:results_discussion}. 

Figure \ref{fig:Example_fits} shows samples of emission-line models of SDSS J092041.76$+$072544.0 and J133254$+$005250, which are both lower redshift for our sample and contain the \hbeta\ emission feature. Models of \civ, \mgii, and \hbeta\ are presented. In this figure, both \mgii\ and \hbeta\ models use an underlying T06 \feii\ template. Examples of all line models of \civ, \mgii, and \hbeta, showing the emission-line models in greater detail, are provided as online supplementary material, where models of \mgii\ and \hbeta\ are separated by \feii\ template. Due to the lack of a continuum redward of \hbeta, we can only use the blue side to constrain the continuum.

\section{Single-Epoch Virial Mass Estimate} \label{sec:virial_mbh}
We measure the black hole mass from single-epoch spectroscopic data using the virial estimate, which is a method routinely applied to QSO spectra \citep[e.g.,][]{Vestergaard_2002, Mclure_2002, Mclure_2004, Greene_2005, Vestergaard_2006}. The model assumes that the motion of gas around the black hole is virialized and its dynamics are dominated by the central gravitational field. The velocity-broadened line profile measures the gas velocity and the nuclear continuum luminosity is used as a proxy for the radius of the BLR. The radius-luminosity ($R$-$L$) relationship is an empirical correlation derived from reverberation mapping experiments which tightly links the radius of the BLR to the continuum luminosity \citep[e.g.,][]{Kaspi_2000, Kaspi_2005, Bentz_2006, Bentz_2013}. Common emission-lines used to estimate gas velocity include \hbeta, \civ, and \mgii, but the \hbeta\ line is redshifted out of the X-shooter NIR coverage at $z \gtrsim 4$. Instead, the \mgii\, emission-line profile is found to generally be correlated with \hbeta\ and can be used as its substitute in single-epoch virial black hole mass estimates \citep[e.g.,][]{Salviander_2007, Shen_2008, Wang_2009, Shen_2012}. Additionally, there are indications that the \mgii-based estimator is more reliable for QSOs with large (> 4000 km s$^{-1}$) \hbeta\ FWHM \citep[see][]{Marziani_2013}. The following equation describes the single-epoch virial mass estimate,
\begin{equation}
   \left(\frac{M_{\rm{BH,vir}}}{M_{\odot}}\right) = 10^{\rm{a}} \left[\frac{\lambda L_{\lambda}}{10^{44} \,\rm{erg\, s^{-1}}}\right]^{b} \left[\frac{\rm{FWHM_{\rm{line}}}}{1000 \,\rm{km\, s^{-1}}}\right]^{2} \,,
   \label{eq:mgii_virial}
\end{equation}
where $\lambda L_{\lambda}$ is the monochromatic luminosity of the QSO continuum, which we measure from the power-law continuum model, and FWHM$_{\rm{line}}$ is the measured line full-width half-maximum of the total broad line profile. We opt to use the FWHM for the virial mass estimate instead of the line dispersion, i.e., the second moment of the line profile. Although the dispersion is well-defined for arbitrary line profiles and may have advantages over the FWHM \citep[e.g.,][]{Fromerth_2000, Peterson_2004, Collin_2006, Rafiee_2011, DallaBonta_2020}, in practice, the line dispersion is sensitive to the wings of the line profile, which are naturally low in flux and can often be difficult to constrain independently from noise due to the accretion disk and \feii\ continuum. We determine all line properties, including the FWHM of \mgii\ and \hbeta, as the average of the measurements obtained from spectral decomposition using each of the all four \feii\ templates. We further quantify the deviation of the measured black hole mass using each template in Appendix \ref{sec:appendix-bhcompare}.

The exponents (a, b) in Equation \ref{eq:mgii_virial} depend on the choice of line and luminosity and are empirically calibrated by reverberation mapping experiments. For the \mgii\ line and a monochromatic luminosity at rest-frame 3000 \AA, (a, b) are calibrated to the values (6.86, 0.5) in \citet{Vestergaard_2009} and (6.74, 0.62) in \citet{Shen_2011}. On average, the differences between these different calibrations are 0.1 dex, but virial mass estimators show an instrinsic scatter of $\sim0.3$ dex around their reverberation mapping counterparts \citep[][]{DallaBonta_2020}, while the reverberation-based estimates exhibit an intrinsic scatter of $\sim0.4$ dex around the $M_{\rm BH}-\sigma_{*}$  relation \citep{Bennert_2021}, meaning the virial mass estimates could have errors as large as $\sim0.5$ dex. We adopt 0.5 dex as our single-epoch virial black hole mass uncertainty in this study. In this study, we use the \mgii-based calibration from \citet{Shen_2011}, which is anchored to a high-luminosity subset of local reverberation mapping determinations from \hbeta, making it better suited to the XQ-100 sources. Other broad emission-lines present in our spectra can be used to obtain virial estimates of the black hole mass as well. Compared with the \mgii\ line, the \civ\ line is more likely to be affected by non-virial motions, such as the radiatively driven outflows \citep[e.g.,][]{Proga_2000, Saturni_2018}, making it potentially a biased black hole mass estimator \citep[e.g.,][]{baskin_laor_2005, Sulentic_2007, Shen_2008}. We provide a measure of the \civ\ blueshift, a signature of outflowing emission \citep{Richards_2011}, in order to quantify how much the \civ-based black hole masses may be biased by non-virial components. The velocity shifts of \civ\ are measured relative to the systemic redshifts from \citet{Lopez_2016_XQ100}.

Table \ref{tab:virial_relations} presents the virial relations and specific calibrations used in this study for determining the black hole mass using \civ, \mgii, and \hbeta\ emission-lines. Using the 3000 \AA\ luminosity, we also estimate the bolometric luminosity by adopting a fixed bolometric correction factor of 5.15, which can lead to errors as large as 50\%, or $\sim$0.3 dex, for individual QSOs \citep{Richards_2006}.

The typical final measurement uncertainties are $\sim$240 km s$^{-1}$ for the \mgii\ FWHM and 0.01 dex for the 3000 \AA\ monochromatic luminosity, resulting in an average of 0.06 dex uncertainty in the \mgii\ black hole mass estimate. Similarly for \hbeta, the average uncertainty is $\sim$640 km s$^{-1}$ for the FWHM and 0.02 dex for the 5100 \AA\ luminosity, resulting in 0.12 dex mean uncertainty in $M_{\rm{BH}}$. Therefore, the measurement uncertainty for both estimates are well below the errors of the virial mass estimator. Without the additional uncertainties introduced by the multiple \feii\ templates, the mean final measurement uncertainty in \civ\ FWHM is $\sim$130 km s$^{-1}$ with negligible uncertainty in the 1450\AA\ luminosity. The typical black hole mass uncertainty from the \civ\ virial estimator is thus 0.02 dex. However, line asymmetries and contribution from QSO outflows or disk winds should imply a greater uncertainty of \civ-based black hole masses.

\begingroup
\begin{table}
\centering{
\caption {\label{tab:virial_relations} Virial relations used in this study (see Equation \ref{eq:mgii_virial})}  
\begin{tabular}{lcccc}
\hline \hline
 Emission-line & Luminosity  & a & b & Ref\\
 \hline
\civ\ & 1450 \AA\ & 6.66 & 0.53 & 1 \\
\mgii\ & 3000 \AA\ & 6.74 & 0.62 & 2 \\
\hbeta\ & 5100 \AA\ & 6.91 & 0.50 & 1 \\
\hline \hline
\multicolumn{5}{l}{\footnotesize
$^{1}$ \citet{Vestergaard_2006}
$^{2}$ \citet{Shen_2011}}
\end{tabular}
}
\end{table}
\endgroup

\section{Results and Discussion} \label{sec:results_discussion}
For each of the 3 broad emission-lines (\civ, \mgii, and \hbeta) used for virial black hole mass estimates in this study, we measure 6 properties of the broad line profile, described in Table \ref{tab:XQ100_line_properties}. The FWHM of each line is used in the virial mass estimate. The line dispersion, Sigma, is the second moment of the line profile. We also measure the Blueshift, equivalent width (EW), and wavelength of the line profile peak (pWavelength). The blueshift is measured from the median wavelength bisecting the total flux of the line profile, and can be a useful indicator of QSO orientation, particularly with the \civ\ line \citep[e.g.,][]{Richards_2002, Yong_2020}. We measure the integrated line luminosity (iLuminosity) from the reconstructed broad emission-line profile. Ratios of the integrated luminosity may be used for chemical abundance estimates \citep[e.g.,][]{Hamann_1999, Hamann_2002, Nagao_2006}, while the EWs may be used in studies of the Baldwin effect \citep[e.g.,][]{Baldwin_1977, Patino_2016}.
We present a sample of measured emission-line properties for 5 selected QSOs in Table \ref{tab:XQ100_properties_sample}, while the full table is available as online supplementary material. 

\begingroup
\begin{table}
\centering{
\caption {\label{tab:XQ100_line_properties} Description of measured properties for each broad emission-line.}  
\begin{tabular}{lll}
\hline \hline
 Suffix & Description  & Units\\
 \hline
FWHM & Full-width half-maximum of profile & km s$^{-1}$\\
Sigma & Second moment of profile & km s$^{-1}$ \\
Blueshift & Defined by the flux-bisecting wavelength &  km s$^{-1}$ \\
EW & Equivalent width in rest-frame & \AA \\
pWavelength & Peak wavelength & \AA \\
iLuminosity & Integrated log luminosity & erg s$^{-1}$\\
\hline \hline
\end{tabular}
}
\end{table}
\endgroup

\subsection{QSO Variability} \label{sec:variability}
Ever since the identification of the first QSOs, it has been recognized that QSOs are intrinsically variable \citep{Matthews_1963}. Variations of QSO brightness occur on a large range of timescales from hours to years, where short timescales are typically associated with higher energy X-ray flux and longer timescales to the disk emission \citep{Edelson_2015, Lira_2015}. Models of QSO variability focus on its stochastic origin, comparing the ensemble variability structure function (SF) to damped random walk (DRW) models \citep[e.g.,][]{Kelly_2009, MacLeod_2010, Kozlowski_2016, Suberlak_2021}. 

Many studies have shown that the amplitude of QSO variability is anti-correlated with the QSO luminosity, with little apparent dependence on the redshift \citep[e.g.,][]{VandenBerk_2004, MacLeod_2010, Kozlowski_2016, Caplar_2017}. For a high-redshift and high-luminosity sample, such as XQ-100, the long-term asymptotic variability amplitude (SF$_{\infty}$) is measured to be low, from 0.1 mag \citep[e.g.,][]{MacLeod_2010, Suberlak_2021} to 0.25 mag \citep[e.g.,][]{Kozlowski_2016}, where these studies made use of SDSS Stripe 82 \citep{Jiang_2014_Stripe82}, an equatorial region imaged repeatedly during 2005, 2006, and 2007.

In this study, our sample of XQ-100 QSOs is flux calibrated to photometry observed at a separate epoch, thus our results are susceptible to QSO variability. In order to constrain variability in the XQ-100 sample, we crossmatch all sources with the Pan-STARRS DR2 detections table \citep{Flewelling_2020}, removing all cases for which the number of multi-epoch \textit{i}-band detections is less than 5. This results in a nearly complete sample of 82 QSOs, each with up to 33 independent \textit{i}-band detections across a period of 3-5 years from 2009-2015. Models of QSO variability characteristic timescales typically find a best-fit parameter of a few hundred days for supermassive black holes in the rest-frame \citep[e.g.,][]{MacLeod_2010, Burke_2021, Suberlak_2021}, so the Pan-STARRS detections cover little more than one rest-frame characteristic timescale. For this analysis, we estimate the observed variability amplitude using the structure function, SF$_{\rm{obs}}(\Delta t)$ = $\rm{rms}\left[m(t) - m(t+\Delta t)\right]$, where rms is the root-mean-square deviation. We calculate SF$_{\rm{obs}}(\Delta t_{\rm{obs}})$ from the ensemble of 82 QSOs with multi-epoch photometric measurements by considering the distribution of $\Delta m$, the measured magnitude difference, for each pair of measurements separated by a time-lag, $\Delta t_{\rm{obs}} \sim$ 2, 3, or 4 years in the observed frame.

Using Pan-STARRS DR2 \textit{i}-band detections, we find the ensemble variability amplitude of the XQ-100 sample to be SF$_{\rm{obs}}(\Delta t_{\rm{obs}})$ = (0.125, 0.132, 0.150) mag for observed frame $\Delta t_{\rm{obs}} \sim$ (2, 3, 4) years, which is consistent with a luminous QSO asymptotic long-term variability of SF$_{\infty} < 0.20$ \citep{MacLeod_2010, Kozlowski_2016}. The photometric calibrations we have used span an even wider timeframe relative to the spectroscopic observations taken between 2012--2014. Therefore, we assume the asymptotic variability as our uncertainty in the overall flux normalisation for each spectrum. A 0.20 mag variability amplitude between the X-shooter and photometric observation would manifest as $< 0.1$ dex uncertainty in the measured luminosities. As a consequence, we expect a black hole mass uncertainty up to $\sim$0.05 dex may be present from variability.

\subsection{XQ-100 Sample Properties} \label{sec:sample-props}
\begin{figure*}
	\includegraphics[width=1.0\textwidth]{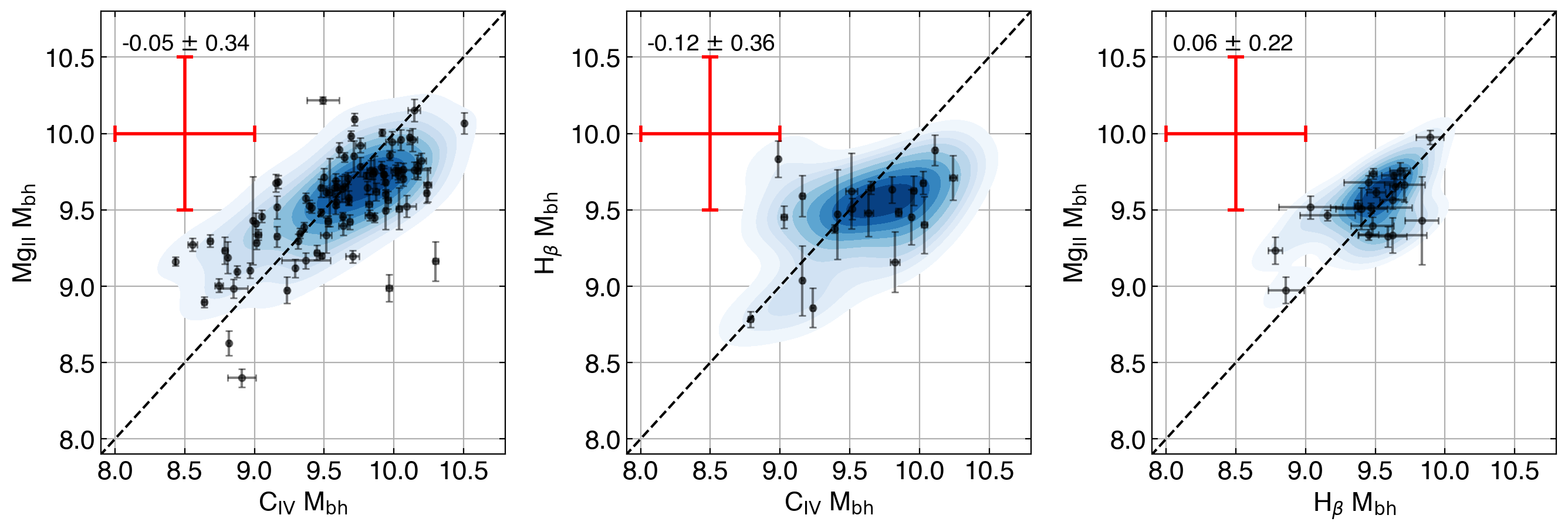}
    \caption{Comparison of virial black hole masses based on \mgii, \civ, and \hbeta\ relations in Table \ref{tab:virial_relations}. The blue shaded contours represent the two-dimensional continuous probability density distribution calculated with a kernel density estimator \citep{Waskom2021_seaborn}. Each subsequent contour level marks density iso-proportions increasing by an additional 10\% up to 90\% enclosed. The red error bars plotted on the top left of each plot show the extent of the 0.5 dex uncertainty, which is a conservative estimate of the uncertainty inherent in the virial mass estimation method. Comparisons between all three virial mass estimates are scatted around the 1:1 relation, indicated by the black dashed line. The mean and standard deviation listed in the top-left of each panel are based on the residual from the mass measure on the y-axis subtracted by the mass measure on the x-axis.}
    \label{fig:mass_compare}
\end{figure*}

We now examine the black hole mass estimates from the \civ, \mgii, and \hbeta-based virial estimators and contextualise the results. For both the \hbeta\ and \mgii\ lines, we subtract the narrow component to obtain the pure broad emission profile \citep[e.g.,][]{Kovacevic_2017} and measure the broad-line properties. Figure \ref{fig:mass_compare} compares the three mass estimates to each other. As we do not exclude any flagged targets from further analysis and contextualisation, the \civ\ and \mgii\ comparison contains all 100 QSOs in the sample and the comparisons to \hbeta\ are limited to 21 measurements. All three panels show data dispersed around the 1:1 relation denoted by the black dashed line, where the total sample variance is smaller than the adopted 0.5 dex uncertainty of the virial mass relation, which is shown for scale on the top-left of each plot. The mean differences and the standard deviation between black hole mass estimates are $\log\left(\rm{M_{\mgii}}/\rm{M_{\civ}}\right) = -0.05 \pm 0.34$, $\log\left(\rm{M_{\hbeta}}/\rm{M_{\civ}}\right) = -0.12 \pm 0.36$, and $\log\left(\rm{M_{\mgii}}/\rm{M_{\hbeta}}\right) = 0.06 \pm 0.22$. In the online supplementary table, we provide an averaged black hole mass estimate from all measured lines and determine a ``Mbh\_Flag'' for when the averaged masses differ from the \mgii-based masses by more than 0.3 dex. Throughout the XQ-100 sample, 8\% of QSOs are flagged in this way, and only one (SDSSJ1042$+$1957) has a \hbeta\ virial mass estimate to shed light on the discrepancy between \civ\ and \mgii-based masses. For SDSSJ1042$+$1957, the \hbeta-based mass estimate is much more consistent with \civ\ than \mgii. The reason may be that the emission-lines are relatively narrow (FWHM$\sim$2000 km s$^{-1}$) compared to the rest of the sample, and only the \mgii\ line models consistently contain a narrow component.

Although the mass measurement for the XQ-100 sample relies on an extrapolation of the well-determined and lower luminosity \hbeta\ reverberation mapping $R$-$L$ relation \citep[e.g.,][]{Bentz_2013}, we find all three virial estimators to remain consistent with each other within the measurement uncertainties in the high luminosity regime. This shows that the relative physical geometry of the three line-emitting regions does not change significantly with luminosity. Additionally, there are minimal systematic differences between our models to individual emission-lines and this increases our confidence in the resulting mass estimate. 

In the case of outliers such as SDSSJ1202--0054, where M$_{\rm{\hbeta}} \sim 9.8$ and M$_{\rm{\civ}} \sim 9.0$, or the inverse scenario for J1320299--052335, where M$_{\rm{\hbeta}} \sim 9.2$ and M$_{\rm{\civ}} \sim 9.8$, their \hbeta\ profiles are truncated and broad components are not well constrained. We present SDSSJ1202--0054 in additional detail in Appendix \ref{sec:appendix-bhcompare}. Other outliers, such as SDSS J074711.15$+$273903.3 which exhibits a > 1 dex mass difference between different lines, are characterised by relatively poor data quality in the wavelength regions surrounding the \mgii\ line, resulting in weaker constraints on the \feii\ continuum model and a narrower emission-line FWHM. Residual telluric features from an insufficient telluric correction can corrupt the continuum model. Additionally, for targets with $3.8 \lesssim z \lesssim 4.2$, the \mgii\ line overlaps with a wide H$_{\rm{2}}$O telluric absorption band at 1.4 $\mu$m which deteriorates the quality of its detection.

\begin{figure*}
	\includegraphics[width=0.8\textwidth]{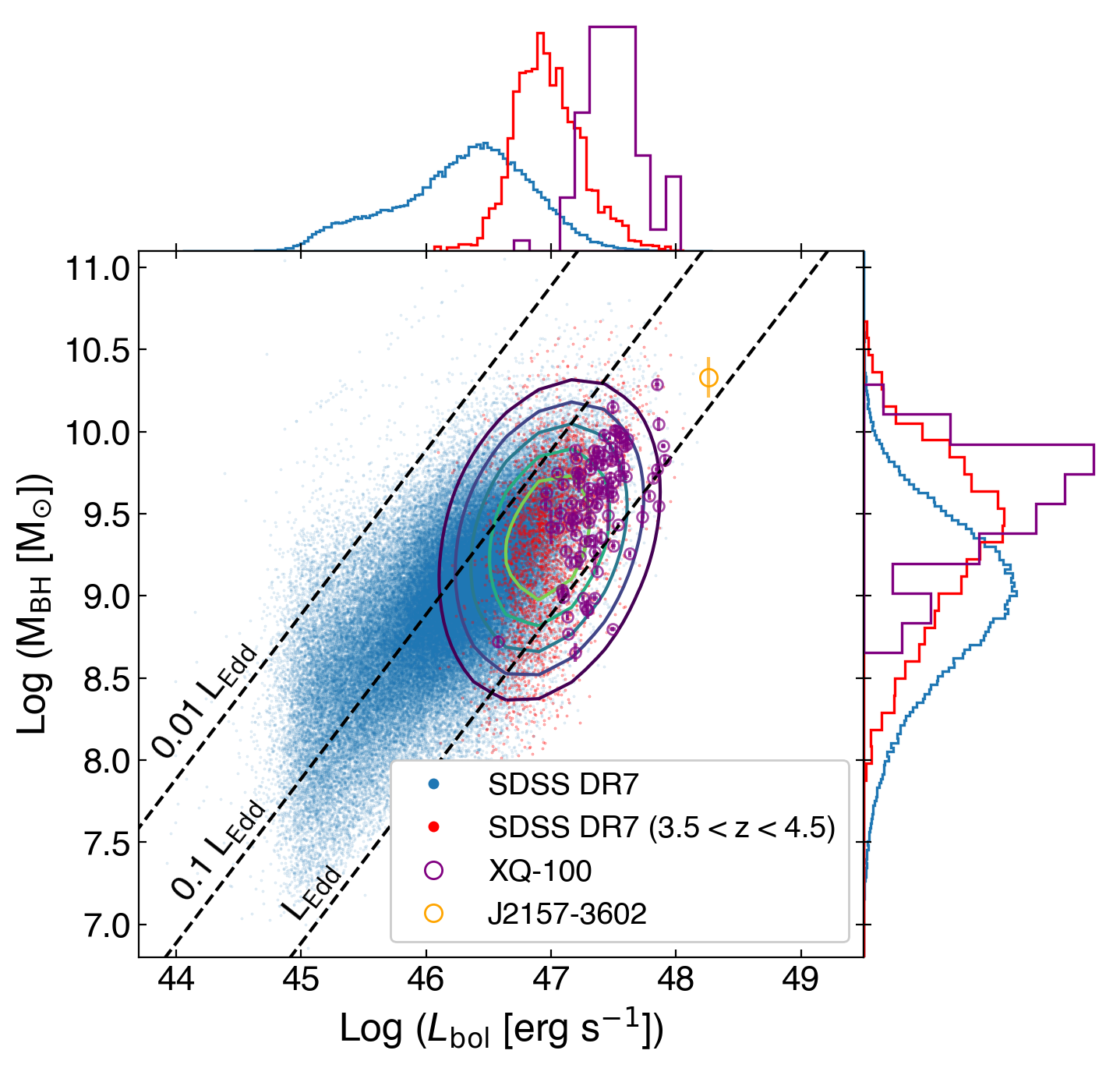}
    \caption{Distribution of XQ-100 black hole masses and luminosities compared to the SDSS DR7 QSO catalogue from \citet{Shen_2011}. The mean virial black hole mass measurements of XQ-100 are shown in purple and the SDSS DR7 data points are shown in blue. The contours delineate iso-proportions in the continuous probability distribution of the higher redshift SDSS sub-sample calculated with a kernel density estimator \citep{Waskom2021_seaborn}. Each contour encloses an additional 10\% up to a 50\% threshold. The black hole mass and bolometric luminosity histograms of the XQ-100 sample are normalised to the same area. Compared to the SDSS DR7 QSO catalogue, the XQ-100 sample occupies the high mass and high luminosity tail. The orange point is J2157$-$3602 ($z = 4.692$), one of the most luminous known QSOs \citep{onken_2020_J2157}.}
    \label{fig:XQ100_Lbol_Mbh}
\end{figure*}

We compare the distribution of black hole masses and bolometric luminosities in XQ-100 to the SDSS DR7 QSO catalogue from \citet{Shen_2011} in Figure \ref{fig:XQ100_Lbol_Mbh}, using mean masses from at least two virial mass estimates to represent the XQ-100 sample. The QSOs in the SDSS DR7 catalogue cover $0.06 < z < 5.47$ in redshift and their black hole masses are primarily based on the \mgii\ emission-line with the same virial mass calibration we have used, but $\sim$40\% of the sample utilise either \civ\ or \hbeta\ with calibrations from \citet{Vestergaard_2006}. The median black hole mass and bolometric luminosity for the SDSS DR7 QSO catalogue is $\log{(\rm{M_{BH}}/\rm{M_\odot})} = 9.0^{+0.5}_{-0.6}$ and $\log{(\rm{L_{bol}}/\rm{erg\,s^{-1}})} = 46.4^{+0.5}_{-0.7}$, where the asymmetric dispersion is set by the 16\% and 84\% percentile.

We also identify the sub-sample of the SDSS DR7 QSO catalogue consisting of 3127 QSOs within the $3.5 < z < 4.5$ redshift range of the XQ-100 sample. Relative to the full catalogue of 104746 objects, the sub-sample has higher median black hole mass and luminosity with $\log{(\rm{M_{BH}}/\rm{M_\odot})} = 9.3^{+0.5}_{-0.8}$ and $\log{(\rm{L_{bol}}/\rm{erg\,s^{-1}})} = 47.0^{+0.3}_{-0.3}$. The XQ-100 sample is more tightly distributed at the high-mass and high-luminosity tail of the redshift-selected SDSS DR7 QSO sub-sample with $\log{(\rm{M_{BH}}/\rm{M_\odot})} = 9.6^{+0.3}_{-0.4}$ and $\log{(\rm{L_{bol}}/\rm{erg\,s^{-1}})} = 47.5^{+0.2}_{-0.2}$. A sub-sample (27\%) of the XQ-100 sample exhibits mildly super-Eddington accretion rates. We also plot J2157$-$3602, one of the most luminous known QSO \citep{onken_2020_J2157}, which is at a comparable redshift ($z = 4.692$), with a black hole mass of $\log{(\rm{M_{BH}}/\rm{M_\odot})} = 10.33$ and bolometric luminosity $\log{(\rm{L_{bol}}/\rm{erg\,s^{-1}})} = 48.4$, measured with the same approach used here \citep{Lai_23_AD}. The full range of XQ-100 QSO properties is measured to span $\log{(\rm{M_{BH}}/\rm{M_\odot})} = 8.6-10.3$ in black hole mass and $\log{(\rm{L_{bol}}/\rm{erg\,s^{-1}})} = 46.7-48.0$ in bolometric luminosity, where over 85\% of the sample lies within $\log(\rm{M_{BH}}/\rm{M_\odot}) = 9-10$ and $\log{(\rm{L_{bol}}/\rm{erg\,s^{-1}})} = 47-48$. 

We find that 55 of the targets have \civ\ measurements in the \citet{Shen_2011} catalogue, which has also produced \civ-based virial mass estimates using the same \citet{Vestergaard_2006} calibration. The mean and standard deviation of differences between the mass estimates from our work and from \citet{Shen_2011} is $\log(\rm{M_{\rm{\civ}}}/\rm{M_{\rm{Shen}}}) = -0.07\pm0.18$ using the \civ-based mass from our sample and $\log(\rm{M_{\rm{avg}}}/\rm{M_{\rm{Shen}}}) = -0.12\pm0.24$ using the mean mass, which is a small systematic adjustment towards lower masses on average. We find no significant correlations between these mass differences and other measurable line properties.

\begingroup
\begin{table*}
\caption {\label{tab:XQ100_properties_sample} Measured emission-line properties of \mgii, \civ, and \hbeta\ and continuum measurements for a selected sample of QSOs. Refer to Table \ref{tab:XQ100_line_properties} for the entry explanations. The reported uncertainties are sourced from the measurement and do not include the additional error of the virial mass method or QSO variability. The full table containing details for all XQ-100 QSOs is available as online supplementary material.}  
\begin{tabular}{llrrrrr}
\hline \hline
& Units & HB89 0000-263 & PMN J0100-2708 & BRI 0241-0146 & J112634-012436 & J1401+0244 \\ 
\hline
OBJECT &  & HB89 0000-263 & PMN J0100-2708 & BRI 0241-0146 & J112634-012436 & J1401+0244 \\ 
RA &  & 00:03:22.79 & 01:00:12.47 & 02:44:01.83 & 11:26:34.42 & 14:01:46.52 \\ 
Dec &  & -26:03:19.40 & -27:08:52.10 & -01:34:06.30 & -01:24:38.00 & 02:44:37.70 \\ 
redshift &  & 4.125 & 3.546 & 4.055 & 3.765 & 4.408 \\ 
Source\_i &  & SkyMapper & SkyMapper & SkyMapper & SkyMapper & SkyMapper \\ 
imag & mag & 17.075 $\pm$ 0.006 & 18.928 $\pm$ 0.026 & 18.099 $\pm$ 0.028 & 19.038 $\pm$ 0.071 & 18.395 $\pm$ 0.014 \\ 
Source\_J &  & VHS & VIKINGDR5 & VHS & UKIDSS & UKIDSS \\ 
Jmag & mag & 16.023 $\pm$ 0.008 & 17.590 $\pm$ 0.011 & 16.916 $\pm$ 0.015 & 18.053 $\pm$ 0.038 & 17.419 $\pm$ 0.032 \\ 
CIV\_FWHM & km s$^{-1}$ & 5275 $\pm$ 34 & 6103 $\pm$ 228 & 8387 $\pm$ 418 & 5746 $\pm$ 64 & 6048 $\pm$ 195 \\ 
CIV\_Sigma & km s$^{-1}$ & 3820 $\pm$ 91 & 2888 $\pm$ 367 & 3568 $\pm$ 229 & 3662 $\pm$ 134 & 3888 $\pm$ 276 \\ 
CIV\_Blueshift & km s$^{-1}$ & 1206 $\pm$ 45 & 1618 $\pm$ 110 & 1833 $\pm$ 228 & 2191 $\pm$ 44 & 821 $\pm$ 97 \\ 
CIV\_EW & \AA & 27.370 $\pm$ 0.270 & 23.930 $\pm$ 1.060 & 23.310 $\pm$ 1.100 & 23.270 $\pm$ 0.490 & 39.040 $\pm$ 0.450 \\ 
CIV\_pWavelength & \AA & 1543.380 $\pm$ 0.140 & 1540.830 $\pm$ 0.510 & 1540.860 $\pm$ 0.740 & 1539.290 $\pm$ 0.670 & 1545.790 $\pm$ 0.770 \\ 
CIV\_iLuminosity & erg s$^{-1}$ & 45.754 $\pm$ 0.004 & 44.833 $\pm$ 0.019 & 45.256 $\pm$ 0.021 & 44.819 $\pm$ 0.008 & 45.430 $\pm$ 0.005 \\ 
CIV\_PL\_slope &  & -1.622 $\pm$ 0.008 & -1.352 $\pm$ 0.020 & -1.525 $\pm$ 0.016 & -1.340 $\pm$ 0.025 & -1.374 $\pm$ 0.008 \\ 
MgII\_FWHM & km s$^{-1}$ & 3396 $\pm$ 111 & 3599 $\pm$ 238 & 6378 $\pm$ 537 & 4574 $\pm$ 288 & 4319 $\pm$ 313 \\ 
MgII\_Sigma & km s$^{-1}$ & 3379 $\pm$ 113 & 2810 $\pm$ 297 & 3488 $\pm$ 386 & 3761 $\pm$ 75 & 3410 $\pm$ 245 \\ 
MgII\_Blueshift & km s$^{-1}$ & 217 $\pm$ 71 & 434 $\pm$ 171 & -239 $\pm$ 100 & -57 $\pm$ 132 & -289 $\pm$ 103 \\ 
MgII\_EW & \AA & 21.620 $\pm$ 1.670 & 28.940 $\pm$ 3.150 & 40.640 $\pm$ 3.270 & 36.500 $\pm$ 2.280 & 37.360 $\pm$ 2.110 \\ 
MgII\_pWavelength & \AA & 2798.540 $\pm$ 2.310 & 2792.860 $\pm$ 0.990 & 2808.220 $\pm$ 4.840 & 2794.430 $\pm$ 0.650 & 2804.110 $\pm$ 0.450 \\ 
MgII\_iLuminosity & erg s$^{-1}$ & 45.210 $\pm$ 0.039 & 44.502 $\pm$ 0.059 & 45.083 $\pm$ 0.038 & 44.502 $\pm$ 0.028 & 44.961 $\pm$ 0.032 \\ 
MgII\_PL\_slope &  & -1.339 $\pm$ 0.087 & -1.145 $\pm$ 0.023 & -1.519 $\pm$ 0.140 & -1.685 $\pm$ 0.087 & -1.400 $\pm$ 0.155 \\ 
Hbeta\_FWHM & km s$^{-1}$ &  & 5308 $\pm$ 924 &  &  &  \\ 
Hbeta\_Sigma & km s$^{-1}$ &  & 3864 $\pm$ 637 &  &  &  \\ 
Hbeta\_Blueshift & km s$^{-1}$ &  & -634 $\pm$ 330 &  &  &  \\ 
Hbeta\_EW & \AA &  & 86.440 $\pm$ 25.040 &  &  &  \\ 
Hbeta\_pWavelength & \AA &  & 4856.340 $\pm$ 4.640 &  &  &  \\ 
Hbeta\_iLuminosity & erg s$^{-1}$ &  & 44.494 $\pm$ 0.099 &  &  &  \\ 
Hbeta\_PL\_slope &  &  & -2.426 $\pm$ 0.395 &  &  &  \\ 
log\_L1450 & erg s$^{-1}$ & 47.522 $\pm$ 0.001 & 46.651 $\pm$ 0.001 & 47.090 $\pm$ 0.001 & 46.647 $\pm$ 0.001 & 47.038 $\pm$ 0.001 \\ 
log\_L3000 & erg s$^{-1}$ & 47.312 $\pm$ 0.018 & 46.486 $\pm$ 0.021 & 46.907 $\pm$ 0.023 & 46.369 $\pm$ 0.013 & 46.824 $\pm$ 0.019 \\ 
log\_L5100 & erg s$^{-1}$ &  & 46.240 $\pm$ 0.057 &  &  &  \\ 
logMBH\_CIV & M$_{\odot}$ & 9.971 $\pm$ 0.006 & 9.636 $\pm$ 0.032 & 10.145 $\pm$ 0.043 & 9.582 $\pm$ 0.010 & 9.833 $\pm$ 0.028 \\ 
CIV\_Quality\_Flag &  & 1 &  &  &  &  \\ 
logMBH\_MgII & M$_{\odot}$ & 9.855 $\pm$ 0.030 & 9.394 $\pm$ 0.059 & 10.152 $\pm$ 0.074 & 9.529 $\pm$ 0.055 & 9.762 $\pm$ 0.064 \\ 
MgII\_Quality\_Flag &  &  &  & 1 &  &  \\ 
logMBH\_Hbeta & M$_{\odot}$ &  & 9.480 $\pm$ 0.154 &  &  &  \\ 
Hbeta\_Quality\_Flag &  &  & 1 &  &  &  \\ 
logMBH\_avg & M$_{\odot}$ & 9.913 $\pm$ 0.015 & 9.503 $\pm$ 0.056 & 10.148 $\pm$ 0.043 & 9.556 $\pm$ 0.028 & 9.797 $\pm$ 0.035 \\ 
Mbh\_Flag &  &  &  &  &  &  \\ 
NIR\_VIS\_Flag &  &  &  &  &  &  \\ 
Hbeta\_Truncation\_Flag &  &  & 1 &  &  &  \\ 
\hline \hline
\end{tabular}
\end{table*}
\endgroup

\section{Summary and Conclusion} \label{sec:conclusion}
Infrared echelle spectroscopic observations of high-redshift QSOs provide an opportunity to investigate their optical and ultraviolet atomic transitions. The XQ-100 legacy survey provides a high-quality sample of 100 QSOs in the redshift range of $z=3.5-4.5$ with high SNR, wide spectroscopic coverage between its three observation arms, and moderate resolving power.

In this study, we examine rest-frame UV and optical broad-emission-lines from all 100 QSOs in the XQ-100 legacy survey. We measure properties of the 
\civ, \mgii, and \hbeta\ emission-lines as well as the QSO continuum to estimate QSO luminosities and black hole masses through virial relations. The main results of this study are as follows:

\begin{itemize}
    \item We measure the \civ\ and \mgii\ line for all 100 QSOs and the \hbeta\ line for 21 QSOs, using multiple templates to estimate the underlying \feii\ emission. The virial mass estimate is based on the measured FWHM of all three broad emission-lines and the continuum luminosity measured near each respective emission-line at 1450, 3000, and 5100 \AA. We provide an averaged black hole mass estimate from all measured emission-lines for each QSO in the online supplementary table$^{\ref{github_footnote}}$ and determine the black hole masses of the XQ-100 sample to be $\log{(\rm{M_{BH}}/\rm{M_\odot})} = 8.6-10.3$. A comparison of mass measurements between the \mgii\ virial mass estimate and the \civ\ and \hbeta\ virial estimates show a mean difference and standard deviation of $-0.05 \pm 0.34$ dex and $0.06 \pm 0.22$ dex, respectively, which are both well below the 0.5 dex uncertainty of the virial estimate. There is a general consistency between the mass estimates derived from the \civ, \mgii, and \hbeta\ broad emission lines. Using a fixed 5.15 bolometric correction factor applied to the 3000 \AA\ continuum luminosity, we estimate the bolometric luminosity range of the XQ-100 sample to be $\log{(\rm{L_{bol}}/\rm{erg\,s^{-1}})} = 46.7-48.0$.
    \item Compared to the SDSS DR7 QSO catalogue, QSOs in the XQ-100 legacy survey occupy the high-mass and high-luminosity tail of the distribution. A sizable sub-sample consisting of 27\% of the XQ-100 QSOs are accreting at mildly super-Eddington rates. 
    \item For each broad emission-line from \civ, \mgii, and \hbeta, we measure 6 properties from the broad line profile and release the full set of measurements as online supplementary material. The measured properties of each line include the full-width half maximum (FWHM), line dispersion, blueshift, equivalent width (EW), wavelength of the peak line profile, and integrated luminosity. We also release example figures of all line models in the sample as online material\footnote{also available at \href{https://github.com/samlaihei/XQ-100}{https://github.com/samlaihei/XQ-100} \label{github_footnote}}.
\end{itemize}

Characterising basic properties of the XQ-100 QSOs enables a variety of follow-up research in QSO astrophysics, from chemical enrichment history using emission-line diagnostics to black hole orientation and QSO outflows. As a sample of some of the most luminous QSOs in redshift $3.5 < z < 4.5$, the XQ-100 targets are among most massive, rapidly accreting black holes in the early universe and likely harbour the most massive and active host galaxies as well. These targets can potentially be used to further investigate the relationship between black holes and their host galaxies in the high redshift universe.

\section*{Acknowledgements}
We thank the anonymous referee for their constructive comments and suggestions which have improved this manuscript. We also thank the authors of \citet{Vestergaard_2001}, \citet{Tsuzuki_2006}, \citet{Bruhweiler_Verner_2008}, \citet{Mejia-Restrepo_2016}, \citet{Boroson_Greene_1992}, and \citet{Park_2022} for producing and sharing their \feii\ emission templates.

S.L. is grateful to the Research School of Astronomy \& Astrophysics at Australian National University for funding his Ph.D. studentship.

CAO was supported by the Australian Research Council (ARC) through Discovery Project DP190100252.

This paper is based on observations made with ESO Telescopes at the La Silla Paranal Observatory under programme ID 189.A-0424. 

The national facility capability for SkyMapper has been funded through ARC LIEF grant LE130100104 from the Australian Research Council, awarded to the University of Sydney, the Australian National University, Swinburne University of Technology, the University of Queensland, the University of Western Australia, the University of Melbourne, Curtin University of Technology, Monash University and the Australian Astronomical Observatory. SkyMapper is owned and operated by The Australian National University's Research School of Astronomy and Astrophysics. The survey data were processed and provided by the SkyMapper Team at ANU. The SkyMapper node of the All-Sky Virtual Observatory (ASVO) is hosted at the National Computational Infrastructure (NCI). Development and support of the SkyMapper node of the ASVO has been funded in part by Astronomy Australia Limited (AAL) and the Australian Government through the Commonwealth's Education Investment Fund (EIF) and National Collaborative Research Infrastructure Strategy (NCRIS), particularly the National eResearch Collaboration Tools and Resources (NeCTAR) and the Australian National Data Service Projects (ANDS).

The Pan-STARRS1 Surveys (PS1) and the PS1 public science archive have been made possible through contributions by the Institute for Astronomy, the University of Hawaii, the Pan-STARRS Project Office, the Max-Planck Society and its participating institutes, the Max Planck Institute for Astronomy, Heidelberg and the Max Planck Institute for Extraterrestrial Physics, Garching, The Johns Hopkins University, Durham University, the University of Edinburgh, the Queen's University Belfast, the Harvard-Smithsonian Center for Astrophysics, the Las Cumbres Observatory Global Telescope Network Incorporated, the National Central University of Taiwan, the Space Telescope Science Institute, the National Aeronautics and Space Administration under Grant No. NNX08AR22G issued through the Planetary Science Division of the NASA Science Mission Directorate, the National Science Foundation Grant No. AST-1238877, the University of Maryland, Eotvos Lorand University (ELTE), the Los Alamos National Laboratory, and the Gordon and Betty Moore Foundation.

The VISTA Hemisphere Survey data products served at Astro Data Lab are based on observations collected at the European Organisation for Astronomical Research in the Southern Hemisphere under ESO programme 179.A-2010, and/or data products created thereof.

This work is based in part on data obtained as part of the UKIRT Infrared Deep Sky Survey and the UKIRT Hemisphere Survey.

This publication has made use of data from the VIKING survey from VISTA at the ESO Paranal Observatory, programme ID 179.A-2004. Data processing has been contributed by the VISTA Data Flow System at CASU, Cambridge and WFAU, Edinburgh.

This publication makes use of data products from the Two Micron All Sky Survey, which is a joint project of the University of Massachusetts and the Infrared Processing and Analysis Center/California Institute of Technology, funded by the National Aeronautics and Space Administration and the National Science Foundation.

Funding for the SDSS and SDSS-II has been provided by the Alfred P. Sloan Foundation, the Participating Institutions, the National Science Foundation, the U.S. Department of Energy, the National Aeronautics and Space Administration, the Japanese Monbukagakusho, the Max Planck Society, and the Higher Education Funding Council for England. The SDSS Web Site is \href{http://www.sdss.org/}{http://www.sdss.org/}.

The SDSS is managed by the Astrophysical Research Consortium for the Participating Institutions. The Participating Institutions are the American Museum of Natural History, Astrophysical Institute Potsdam, University of Basel, University of Cambridge, Case Western Reserve University, University of Chicago, Drexel University, Fermilab, the Institute for Advanced Study, the Japan Participation Group, Johns Hopkins University, the Joint Institute for Nuclear Astrophysics, the Kavli Institute for Particle Astrophysics and Cosmology, the Korean Scientist Group, the Chinese Academy of Sciences (LAMOST), Los Alamos National Laboratory, the Max-Planck-Institute for Astronomy (MPIA), the Max-Planck-Institute for Astrophysics (MPA), New Mexico State University, Ohio State University, University of Pittsburgh, University of Portsmouth, Princeton University, the United States Naval Observatory, and the University of Washington.

This project used public archival data from the Dark Energy Survey (DES). Funding for the DES Projects has been provided by the U.S. Department of Energy, the U.S. National Science Foundation, the Ministry of Science and Education of Spain, the Science and Technology Facilities Council of the United Kingdom, the Higher Education Funding Council for England, the National Center for Supercomputing Applications at the University of Illinois at Urbana-Champaign, the Kavli Institute of Cosmological Physics at the University of Chicago, the Center for Cosmology and Astro-Particle Physics at the Ohio State University, the Mitchell Institute for Fundamental Physics and Astronomy at Texas A\&M University, Financiadora de Estudos e Projetos, Funda{\c c}{\~a}o Carlos Chagas Filho de Amparo {\`a} Pesquisa do Estado do Rio de Janeiro, Conselho Nacional de Desenvolvimento Cient{\'i}fico e Tecnol{\'o}gico and the Minist{\'e}rio da Ci{\^e}ncia, Tecnologia e Inova{\c c}{\~a}o, the Deutsche Forschungsgemeinschaft, and the Collaborating Institutions in the Dark Energy Survey.

The Collaborating Institutions are Argonne National Laboratory, the University of California at Santa Cruz, the University of Cambridge, Centro de Investigaciones Energ{\'e}ticas, Medioambientales y Tecnol{\'o}gicas-Madrid, the University of Chicago, University College London, the DES-Brazil Consortium, the University of Edinburgh, the Eidgen{\"o}ssische Technische Hochschule (ETH) Z{\"u}rich,  Fermi National Accelerator Laboratory, the University of Illinois at Urbana-Champaign, the Institut de Ci{\`e}ncies de l'Espai (IEEC/CSIC), the Institut de F{\'i}sica d'Altes Energies, Lawrence Berkeley National Laboratory, the Ludwig-Maximilians Universit{\"a}t M{\"u}nchen and the associated Excellence Cluster Universe, the University of Michigan, the National Optical Astronomy Observatory, the University of Nottingham, The Ohio State University, the OzDES Membership Consortium, the University of Pennsylvania, the University of Portsmouth, SLAC National Accelerator Laboratory, Stanford University, the University of Sussex, and Texas A\&M University.

Based in part on observations at Cerro Tololo Inter-American Observatory, National Optical Astronomy Observatory, which is operated by the Association of Universities for Research in Astronomy (AURA) under a cooperative agreement with the National Science Foundation.

Software packages used in this study include \texttt{Numpy} \citep{Numpy_2011}, \texttt{Scipy} \citep{Scipy_2020}, \texttt{Astropy} \citep{Astropy_2013}, \texttt{Specutils} \citep{specutils_2022}, \texttt{Matplotlib} \citep{Matplotlib_2007}, and \texttt{seaborn} \citep{Waskom2021_seaborn}.

%%%%%%%%%%%%%%%%%%%%%%%%%%%%%%%%%%%%%%%%%%%%%%%%%%
\section*{Data Availability}
The data underlying this article will be shared on reasonable request to the corresponding author. The post-processed spectra, supplementary table, and figures can be downloaded from a GitHub repository: \href{https://github.com/samlaihei/XQ-100}{https://github.com/samlaihei/XQ-100}.



\bibliographystyle{mnras}
\bibliography{bibliography} % if your bibtex file is called example.bib

\begin{thebibliography}{}
\makeatletter
\relax
\def\mn@urlcharsother{\let\do\@makeother \do\$\do\&\do\#\do\^\do\_\do\%\do\~}
\def\mn@doi{\begingroup\mn@urlcharsother \@ifnextchar [ {\mn@doi@}
  {\mn@doi@[]}}
\def\mn@doi@[#1]#2{\def\@tempa{#1}\ifx\@tempa\@empty \href
  {http://dx.doi.org/#2} {doi:#2}\else \href {http://dx.doi.org/#2} {#1}\fi
  \endgroup}
\def\mn@eprint#1#2{\mn@eprint@#1:#2::\@nil}
\def\mn@eprint@arXiv#1{\href {http://arxiv.org/abs/#1} {{\tt arXiv:#1}}}
\def\mn@eprint@dblp#1{\href {http://dblp.uni-trier.de/rec/bibtex/#1.xml}
  {dblp:#1}}
\def\mn@eprint@#1:#2:#3:#4\@nil{\def\@tempa {#1}\def\@tempb {#2}\def\@tempc
  {#3}\ifx \@tempc \@empty \let \@tempc \@tempb \let \@tempb \@tempa \fi \ifx
  \@tempb \@empty \def\@tempb {arXiv}\fi \@ifundefined
  {mn@eprint@\@tempb}{\@tempb:\@tempc}{\expandafter \expandafter \csname
  mn@eprint@\@tempb\endcsname \expandafter{\@tempc}}}

\bibitem[\protect\citeauthoryear{{Abbott} et~al.,}{{Abbott}
  et~al.}{2021}]{DES_2021}
{Abbott} T.~M.~C.,  et~al., 2021, \mn@doi [\apjs] {10.3847/1538-4365/ac00b3},
  \href {https://ui.adsabs.harvard.edu/abs/2021ApJS..255...20A} {255, 20}

\bibitem[\protect\citeauthoryear{{Astropy Collaboration} et~al.,}{{Astropy
  Collaboration} et~al.}{2013}]{Astropy_2013}
{Astropy Collaboration} et~al., 2013, \mn@doi [\aap]
  {10.1051/0004-6361/201322068}, \href
  {https://ui.adsabs.harvard.edu/abs/2013A&A...558A..33A} {558, A33}

\bibitem[\protect\citeauthoryear{{Baldwin}}{{Baldwin}}{1977}]{Baldwin_1977}
{Baldwin} J.~A.,  1977, \mn@doi [\apj] {10.1086/155294}, \href
  {https://ui.adsabs.harvard.edu/abs/1977ApJ...214..679B} {214, 679}

\bibitem[\protect\citeauthoryear{{Baskin} \& {Laor}}{{Baskin} \&
  {Laor}}{2005}]{baskin_laor_2005}
{Baskin} A.,  {Laor} A.,  2005, \mn@doi [\mnras]
  {10.1111/j.1365-2966.2004.08525.x}, \href
  {https://ui.adsabs.harvard.edu/abs/2005MNRAS.356.1029B} {356, 1029}

\bibitem[\protect\citeauthoryear{{Becker}, {Sargent}, {Rauch}  \&
  {Carswell}}{{Becker} et~al.}{2012}]{Becker_2012}
{Becker} G.~D.,  {Sargent} W. L.~W.,  {Rauch} M.,   {Carswell} R.~F.,  2012,
  \mn@doi [\apj] {10.1088/0004-637X/744/2/91}, \href
  {https://ui.adsabs.harvard.edu/abs/2012ApJ...744...91B} {744, 91}

\bibitem[\protect\citeauthoryear{{Bennert} et~al.,}{{Bennert}
  et~al.}{2021}]{Bennert_2021}
{Bennert} V.~N.,  et~al., 2021, \mn@doi [\apj] {10.3847/1538-4357/ac151a},
  \href {https://ui.adsabs.harvard.edu/abs/2021ApJ...921...36B} {921, 36}

\bibitem[\protect\citeauthoryear{{Bentz}, {Peterson}, {Pogge}, {Vestergaard}
  \& {Onken}}{{Bentz} et~al.}{2006}]{Bentz_2006}
{Bentz} M.~C.,  {Peterson} B.~M.,  {Pogge} R.~W.,  {Vestergaard} M.,   {Onken}
  C.~A.,  2006, \mn@doi [\apj] {10.1086/503537}, \href
  {https://ui.adsabs.harvard.edu/abs/2006ApJ...644..133B} {644, 133}

\bibitem[\protect\citeauthoryear{{Bentz} et~al.,}{{Bentz}
  et~al.}{2013}]{Bentz_2013}
{Bentz} M.~C.,  et~al., 2013, \mn@doi [\apj] {10.1088/0004-637X/767/2/149},
  \href {https://ui.adsabs.harvard.edu/abs/2013ApJ...767..149B} {767, 149}

\bibitem[\protect\citeauthoryear{{Berg} et~al.,}{{Berg}
  et~al.}{2016}]{Berg_2016}
{Berg} T.~A.~M.,  et~al., 2016, \mn@doi [\mnras] {10.1093/mnras/stw2232}, \href
  {https://ui.adsabs.harvard.edu/abs/2016MNRAS.463.3021B} {463, 3021}

\bibitem[\protect\citeauthoryear{{Berg} et~al.,}{{Berg}
  et~al.}{2019}]{Berg_2019}
{Berg} T. A.~M.,  et~al., 2019, \mn@doi [\mnras] {10.1093/mnras/stz2012}, \href
  {https://ui.adsabs.harvard.edu/abs/2019MNRAS.488.4356B} {488, 4356}

\bibitem[\protect\citeauthoryear{{Berg} et~al.,}{{Berg}
  et~al.}{2021}]{Berg_2021}
{Berg} T. A.~M.,  et~al., 2021, \mn@doi [\mnras] {10.1093/mnras/stab184}, \href
  {https://ui.adsabs.harvard.edu/abs/2021MNRAS.502.4009B} {502, 4009}

\bibitem[\protect\citeauthoryear{{Boroson} \& {Green}}{{Boroson} \&
  {Green}}{1992}]{Boroson_Greene_1992}
{Boroson} T.~A.,  {Green} R.~F.,  1992, \mn@doi [\apjs] {10.1086/191661}, \href
  {https://ui.adsabs.harvard.edu/abs/1992ApJS...80..109B} {80, 109}

\bibitem[\protect\citeauthoryear{{Bruhweiler} \& {Verner}}{{Bruhweiler} \&
  {Verner}}{2008}]{Bruhweiler_Verner_2008}
{Bruhweiler} F.,  {Verner} E.,  2008, \mn@doi [\apj] {10.1086/525557}, \href
  {https://ui.adsabs.harvard.edu/abs/2008ApJ...675...83B} {675, 83}

\bibitem[\protect\citeauthoryear{{Burke} et~al.,}{{Burke}
  et~al.}{2021}]{Burke_2021}
{Burke} C.~J.,  et~al., 2021, \mn@doi [Science] {10.1126/science.abg9933},
  \href {https://ui.adsabs.harvard.edu/abs/2021Sci...373..789B} {373, 789}

\bibitem[\protect\citeauthoryear{{Caplar}, {Lilly}  \& {Trakhtenbrot}}{{Caplar}
  et~al.}{2017}]{Caplar_2017}
{Caplar} N.,  {Lilly} S.~J.,   {Trakhtenbrot} B.,  2017, \mn@doi [\apj]
  {10.3847/1538-4357/834/2/111}, \href
  {https://ui.adsabs.harvard.edu/abs/2017ApJ...834..111C} {834, 111}

\bibitem[\protect\citeauthoryear{{Carnall}}{{Carnall}}{2017}]{Carnall_2017}
{Carnall} A.~C.,  2017, arXiv e-prints, \href
  {https://ui.adsabs.harvard.edu/abs/2017arXiv170505165C} {p. arXiv:1705.05165}

\bibitem[\protect\citeauthoryear{{Chambers} et~al.,}{{Chambers}
  et~al.}{2016}]{PanSTARRS}
{Chambers} K.~C.,  et~al., 2016, arXiv e-prints, \href
  {https://ui.adsabs.harvard.edu/abs/2016arXiv161205560C} {p. arXiv:1612.05560}

\bibitem[\protect\citeauthoryear{{Coatman}, {Hewett}, {Banerji}, {Richards},
  {Hennawi}  \& {Prochaska}}{{Coatman} et~al.}{2019}]{Coatman_2019}
{Coatman} L.,  {Hewett} P.~C.,  {Banerji} M.,  {Richards} G.~T.,  {Hennawi}
  J.~F.,   {Prochaska} J.~X.,  2019, \mn@doi [\mnras] {10.1093/mnras/stz1167},
  \href {https://ui.adsabs.harvard.edu/abs/2019MNRAS.486.5335C} {486, 5335}

\bibitem[\protect\citeauthoryear{{Collin}, {Kawaguchi}, {Peterson}  \&
  {Vestergaard}}{{Collin} et~al.}{2006}]{Collin_2006}
{Collin} S.,  {Kawaguchi} T.,  {Peterson} B.~M.,   {Vestergaard} M.,  2006,
  \mn@doi [\aap] {10.1051/0004-6361:20064878}, \href
  {https://ui.adsabs.harvard.edu/abs/2006A&A...456...75C} {456, 75}

\bibitem[\protect\citeauthoryear{{Croton} et~al.,}{{Croton}
  et~al.}{2006}]{Croton_2006}
{Croton} D.~J.,  et~al., 2006, \mn@doi [\mnras]
  {10.1111/j.1365-2966.2005.09675.x}, \href
  {https://ui.adsabs.harvard.edu/abs/2006MNRAS.365...11C} {365, 11}

\bibitem[\protect\citeauthoryear{{D'Odorico} et~al.,}{{D'Odorico}
  et~al.}{2023}]{Dodorico_2023_XQR30}
{D'Odorico} V.,  et~al., 2023, \mn@doi [\mnras] {10.1093/mnras/stad1468}, \href
  {https://ui.adsabs.harvard.edu/abs/2023MNRAS.523.1399D} {523, 1399}

\bibitem[\protect\citeauthoryear{{Dalla Bont{\`a}} et~al.,}{{Dalla Bont{\`a}}
  et~al.}{2020}]{DallaBonta_2020}
{Dalla Bont{\`a}} E.,  et~al., 2020, \mn@doi [\apj] {10.3847/1538-4357/abbc1c},
  \href {https://ui.adsabs.harvard.edu/abs/2020ApJ...903..112D} {903, 112}

\bibitem[\protect\citeauthoryear{{Dietrich}, {Appenzeller}, {Vestergaard}  \&
  {Wagner}}{{Dietrich} et~al.}{2002}]{Dietrich_2002}
{Dietrich} M.,  {Appenzeller} I.,  {Vestergaard} M.,   {Wagner} S.~J.,  2002,
  \mn@doi [\apj] {10.1086/324337}, \href
  {https://ui.adsabs.harvard.edu/abs/2002ApJ...564..581D} {564, 581}

\bibitem[\protect\citeauthoryear{{Dye} et~al.,}{{Dye} et~al.}{2018}]{UHS_DR1}
{Dye} S.,  et~al., 2018, \mn@doi [\mnras] {10.1093/mnras/stx2622}, \href
  {https://ui.adsabs.harvard.edu/abs/2018MNRAS.473.5113D} {473, 5113}

\bibitem[\protect\citeauthoryear{Earl et~al.,}{Earl
  et~al.}{2022}]{specutils_2022}
Earl N.,  et~al., 2022, astropy/specutils: V1.7.0,
  \mn@doi{10.5281/zenodo.6207491}, \url
  {https://doi.org/10.5281/zenodo.6207491}

\bibitem[\protect\citeauthoryear{{Edelson} et~al.,}{{Edelson}
  et~al.}{2015}]{Edelson_2015}
{Edelson} R.,  et~al., 2015, \mn@doi [\apj] {10.1088/0004-637X/806/1/129},
  \href {https://ui.adsabs.harvard.edu/abs/2015ApJ...806..129E} {806, 129}

\bibitem[\protect\citeauthoryear{{Edge}, {Sutherland}, {Kuijken}, {Driver},
  {McMahon}, {Eales}  \& {Emerson}}{{Edge} et~al.}{2013}]{VIKING}
{Edge} A.,  {Sutherland} W.,  {Kuijken} K.,  {Driver} S.,  {McMahon} R.,
  {Eales} S.,   {Emerson} J.~P.,  2013, The Messenger, \href
  {https://ui.adsabs.harvard.edu/abs/2013Msngr.154...32E} {154, 32}

\bibitem[\protect\citeauthoryear{{Flesch}}{{Flesch}}{2015}]{Flesch_2015_HMQ}
{Flesch} E.~W.,  2015, \mn@doi [\pasa] {10.1017/pasa.2015.10}, \href
  {https://ui.adsabs.harvard.edu/abs/2015PASA...32...10F} {32, e010}

\bibitem[\protect\citeauthoryear{{Flewelling} et~al.,}{{Flewelling}
  et~al.}{2020}]{Flewelling_2020}
{Flewelling} H.~A.,  et~al., 2020, \mn@doi [\apjs] {10.3847/1538-4365/abb82d},
  \href {https://ui.adsabs.harvard.edu/abs/2020ApJS..251....7F} {251, 7}

\bibitem[\protect\citeauthoryear{{Fromerth} \& {Melia}}{{Fromerth} \&
  {Melia}}{2000}]{Fromerth_2000}
{Fromerth} M.~J.,  {Melia} F.,  2000, \mn@doi [\apj] {10.1086/308671}, \href
  {https://ui.adsabs.harvard.edu/abs/2000ApJ...533..172F} {533, 172}

\bibitem[\protect\citeauthoryear{{Gaia Collaboration} et~al.,}{{Gaia
  Collaboration} et~al.}{2021}]{GaiaEDR3_2021}
{Gaia Collaboration} et~al., 2021, \mn@doi [\aap]
  {10.1051/0004-6361/202039657}, \href
  {https://ui.adsabs.harvard.edu/abs/2021A&A...649A...1G} {649, A1}

\bibitem[\protect\citeauthoryear{{Grandi}}{{Grandi}}{1982}]{Grandi_1982}
{Grandi} S.~A.,  1982, \mn@doi [\apj] {10.1086/159799}, \href
  {https://ui.adsabs.harvard.edu/abs/1982ApJ...255...25G} {255, 25}

\bibitem[\protect\citeauthoryear{{Greene} \& {Ho}}{{Greene} \&
  {Ho}}{2005}]{Greene_2005}
{Greene} J.~E.,  {Ho} L.~C.,  2005, \mn@doi [\apj] {10.1086/431897}, \href
  {https://ui.adsabs.harvard.edu/abs/2005ApJ...630..122G} {630, 122}

\bibitem[\protect\citeauthoryear{{Greene} et~al.,}{{Greene}
  et~al.}{2010}]{Greene_2010}
{Greene} J.~E.,  et~al., 2010, \mn@doi [\apj] {10.1088/0004-637X/721/1/26},
  \href {https://ui.adsabs.harvard.edu/abs/2010ApJ...721...26G} {721, 26}

\bibitem[\protect\citeauthoryear{{Guo}, {Shen}  \& {Wang}}{{Guo}
  et~al.}{2018}]{Guo_2018}
{Guo} H.,  {Shen} Y.,   {Wang} S.,  2018, {PyQSOFit: Python code to fit the
  spectrum of quasars}, Astrophysics Source Code Library, record ascl:1809.008
  (\mn@eprint {ascl} {1809.008})

\bibitem[\protect\citeauthoryear{{Hamann} \& {Ferland}}{{Hamann} \&
  {Ferland}}{1999}]{Hamann_1999}
{Hamann} F.,  {Ferland} G.,  1999, \mn@doi [\araa]
  {10.1146/annurev.astro.37.1.487}, \href
  {https://ui.adsabs.harvard.edu/abs/1999ARA&A..37..487H} {37, 487}

\bibitem[\protect\citeauthoryear{{Hamann}, {Korista}, {Ferland}, {Warner}  \&
  {Baldwin}}{{Hamann} et~al.}{2002}]{Hamann_2002}
{Hamann} F.,  {Korista} K.~T.,  {Ferland} G.~J.,  {Warner} C.,   {Baldwin} J.,
  2002, \mn@doi [\apj] {10.1086/324289}, \href
  {https://ui.adsabs.harvard.edu/abs/2002ApJ...564..592H} {564, 592}

\bibitem[\protect\citeauthoryear{{H{\"a}ring} \& {Rix}}{{H{\"a}ring} \&
  {Rix}}{2004}]{Haring_2004}
{H{\"a}ring} N.,  {Rix} H.-W.,  2004, \mn@doi [\apjl] {10.1086/383567}, \href
  {https://ui.adsabs.harvard.edu/abs/2004ApJ...604L..89H} {604, L89}

\bibitem[\protect\citeauthoryear{{Hunter}}{{Hunter}}{2007}]{Matplotlib_2007}
{Hunter} J.~D.,  2007, \mn@doi [Computing in Science and Engineering]
  {10.1109/MCSE.2007.55}, \href
  {https://ui.adsabs.harvard.edu/abs/2007CSE.....9...90H} {9, 90}

\bibitem[\protect\citeauthoryear{{Ir{\v{s}}i{\v{c}}}
  et~al.,}{{Ir{\v{s}}i{\v{c}}} et~al.}{2017}]{Irsic_2017}
{Ir{\v{s}}i{\v{c}}} V.,  et~al., 2017, \mn@doi [\mnras]
  {10.1093/mnras/stw3372}, \href
  {https://ui.adsabs.harvard.edu/abs/2017MNRAS.466.4332I} {466, 4332}

\bibitem[\protect\citeauthoryear{{Jiang} et~al.,}{{Jiang}
  et~al.}{2014}]{Jiang_2014_Stripe82}
{Jiang} L.,  et~al., 2014, \mn@doi [\apjs] {10.1088/0067-0049/213/1/12}, \href
  {https://ui.adsabs.harvard.edu/abs/2014ApJS..213...12J} {213, 12}

\bibitem[\protect\citeauthoryear{{Jones}, {Noll}, {Kausch}, {Szyszka}  \&
  {Kimeswenger}}{{Jones} et~al.}{2013}]{Jones_2013}
{Jones} A.,  {Noll} S.,  {Kausch} W.,  {Szyszka} C.,   {Kimeswenger} S.,  2013,
  \mn@doi [\aap] {10.1051/0004-6361/201322433}, \href
  {https://ui.adsabs.harvard.edu/abs/2013A&A...560A..91J} {560, A91}

\bibitem[\protect\citeauthoryear{{Kaspi}, {Smith}, {Netzer}, {Maoz}, {Jannuzi}
  \& {Giveon}}{{Kaspi} et~al.}{2000}]{Kaspi_2000}
{Kaspi} S.,  {Smith} P.~S.,  {Netzer} H.,  {Maoz} D.,  {Jannuzi} B.~T.,
  {Giveon} U.,  2000, \mn@doi [\apj] {10.1086/308704}, \href
  {https://ui.adsabs.harvard.edu/abs/2000ApJ...533..631K} {533, 631}

\bibitem[\protect\citeauthoryear{{Kaspi}, {Maoz}, {Netzer}, {Peterson},
  {Vestergaard}  \& {Jannuzi}}{{Kaspi} et~al.}{2005}]{Kaspi_2005}
{Kaspi} S.,  {Maoz} D.,  {Netzer} H.,  {Peterson} B.~M.,  {Vestergaard} M.,
  {Jannuzi} B.~T.,  2005, \mn@doi [\apj] {10.1086/431275}, \href
  {https://ui.adsabs.harvard.edu/abs/2005ApJ...629...61K} {629, 61}

\bibitem[\protect\citeauthoryear{{Kelly}, {Bechtold}  \&
  {Siemiginowska}}{{Kelly} et~al.}{2009}]{Kelly_2009}
{Kelly} B.~C.,  {Bechtold} J.,   {Siemiginowska} A.,  2009, \mn@doi [\apj]
  {10.1088/0004-637X/698/1/895}, \href
  {https://ui.adsabs.harvard.edu/abs/2009ApJ...698..895K} {698, 895}

\bibitem[\protect\citeauthoryear{{Kelson}}{{Kelson}}{2003}]{Kelson_2003}
{Kelson} D.~D.,  2003, \mn@doi [\pasp] {10.1086/375502}, \href
  {https://ui.adsabs.harvard.edu/abs/2003PASP..115..688K} {115, 688}

\bibitem[\protect\citeauthoryear{{Kova{\v{c}}evi{\'c}-Doj{\v{c}}inovi{\'c}},
  {Mar{\v{c}}eta-Mandi{\'c}}  \&
  {Popovi{\'c}}}{{Kova{\v{c}}evi{\'c}-Doj{\v{c}}inovi{\'c}}
  et~al.}{2017}]{Kovacevic_2017}
{Kova{\v{c}}evi{\'c}-Doj{\v{c}}inovi{\'c}} J.,  {Mar{\v{c}}eta-Mandi{\'c}} S.,
   {Popovi{\'c}} L.~{\v{C}}.,  2017, \mn@doi [Frontiers in Astronomy and Space
  Sciences] {10.3389/fspas.2017.00007}, \href
  {https://ui.adsabs.harvard.edu/abs/2017FrASS...4....7K} {4, 7}

\bibitem[\protect\citeauthoryear{{Kova{\v{c}}evi{\'c}}, {Popovi{\'c}}  \&
  {Kollatschny}}{{Kova{\v{c}}evi{\'c}} et~al.}{2014}]{Kovacevic_2014}
{Kova{\v{c}}evi{\'c}} J.,  {Popovi{\'c}} L.~{\v{C}}.,   {Kollatschny} W.,
  2014, \mn@doi [Advances in Space Research] {10.1016/j.asr.2013.11.035}, \href
  {https://ui.adsabs.harvard.edu/abs/2014AdSpR..54.1347K} {54, 1347}

\bibitem[\protect\citeauthoryear{{Koz{\l}owski}}{{Koz{\l}owski}}{2016}]{Kozlowski_2016}
{Koz{\l}owski} S.,  2016, \mn@doi [\apj] {10.3847/0004-637X/826/2/118}, \href
  {https://ui.adsabs.harvard.edu/abs/2016ApJ...826..118K} {826, 118}

\bibitem[\protect\citeauthoryear{Lai}{Lai}{2023}]{PyQSpecFit_v1}
Lai S.,  2023, samlaihei/PyQSpecFit: PyQSpecFit v1.0.0,
  \mn@doi{10.5281/zenodo.7772752}, \url
  {https://doi.org/10.5281/zenodo.7772752}

\bibitem[\protect\citeauthoryear{{Lai} et~al.,}{{Lai} et~al.}{2022}]{Lai_2022}
{Lai} S.,  et~al., 2022, \mn@doi [\mnras] {10.1093/mnras/stac1001}, \href
  {https://ui.adsabs.harvard.edu/abs/2022MNRAS.513.1801L} {513, 1801}

\bibitem[\protect\citeauthoryear{Lai, Wolf, Onken  \& Bian}{Lai
  et~al.}{2023}]{Lai_23_AD}
Lai S.,  Wolf C.,  Onken C.~A.,   Bian F.,  2023, Characterising SMSS
  J2157-3602, the most luminous known quasar, with accretion disc models,
  submitted

\bibitem[\protect\citeauthoryear{{Lawrence} et~al.,}{{Lawrence}
  et~al.}{2007}]{UKIDSS}
{Lawrence} A.,  et~al., 2007, \mn@doi [\mnras]
  {10.1111/j.1365-2966.2007.12040.x}, \href
  {https://ui.adsabs.harvard.edu/abs/2007MNRAS.379.1599L} {379, 1599}

\bibitem[\protect\citeauthoryear{{Lira}, {Ar{\'e}valo}, {Uttley}, {McHardy}  \&
  {Videla}}{{Lira} et~al.}{2015}]{Lira_2015}
{Lira} P.,  {Ar{\'e}valo} P.,  {Uttley} P.,  {McHardy} I.~M.~M.,   {Videla} L.,
   2015, \mn@doi [\mnras] {10.1093/mnras/stv1945}, \href
  {https://ui.adsabs.harvard.edu/abs/2015MNRAS.454..368L} {454, 368}

\bibitem[\protect\citeauthoryear{{L{\'o}pez} et~al.,}{{L{\'o}pez}
  et~al.}{2016}]{Lopez_2016_XQ100}
{L{\'o}pez} S.,  et~al., 2016, \mn@doi [\aap] {10.1051/0004-6361/201628161},
  \href {https://ui.adsabs.harvard.edu/abs/2016A&A...594A..91L} {594, A91}

\bibitem[\protect\citeauthoryear{{Lyke} et~al.,}{{Lyke}
  et~al.}{2020}]{Lyke_2020}
{Lyke} B.~W.,  et~al., 2020, \mn@doi [\apjs] {10.3847/1538-4365/aba623}, \href
  {https://ui.adsabs.harvard.edu/abs/2020ApJS..250....8L} {250, 8}

\bibitem[\protect\citeauthoryear{{MacLeod} et~al.,}{{MacLeod}
  et~al.}{2010}]{MacLeod_2010}
{MacLeod} C.~L.,  et~al., 2010, \mn@doi [\apj] {10.1088/0004-637X/721/2/1014},
  \href {https://ui.adsabs.harvard.edu/abs/2010ApJ...721.1014M} {721, 1014}

\bibitem[\protect\citeauthoryear{{Marconi} \& {Hunt}}{{Marconi} \&
  {Hunt}}{2003}]{Marconi_2003}
{Marconi} A.,  {Hunt} L.~K.,  2003, \mn@doi [\apjl] {10.1086/375804}, \href
  {https://ui.adsabs.harvard.edu/abs/2003ApJ...589L..21M} {589, L21}

\bibitem[\protect\citeauthoryear{{Marziani}, {Sulentic}, {Plauchu-Frayn}  \&
  {del Olmo}}{{Marziani} et~al.}{2013}]{Marziani_2013}
{Marziani} P.,  {Sulentic} J.~W.,  {Plauchu-Frayn} I.,   {del Olmo} A.,  2013,
  \mn@doi [\aap] {10.1051/0004-6361/201321374}, \href
  {https://ui.adsabs.harvard.edu/abs/2013A&A...555A..89M} {555, A89}

\bibitem[\protect\citeauthoryear{{Matejek} \& {Simcoe}}{{Matejek} \&
  {Simcoe}}{2012}]{Matejek_2012}
{Matejek} M.~S.,  {Simcoe} R.~A.,  2012, \mn@doi [\apj]
  {10.1088/0004-637X/761/2/112}, \href
  {https://ui.adsabs.harvard.edu/abs/2012ApJ...761..112M} {761, 112}

\bibitem[\protect\citeauthoryear{{Matthews} \& {Sandage}}{{Matthews} \&
  {Sandage}}{1963}]{Matthews_1963}
{Matthews} T.~A.,  {Sandage} A.~R.,  1963, \mn@doi [\apj] {10.1086/147615},
  \href {https://ui.adsabs.harvard.edu/abs/1963ApJ...138...30M} {138, 30}

\bibitem[\protect\citeauthoryear{{McConnell} \& {Ma}}{{McConnell} \&
  {Ma}}{2013}]{McConnell_2013}
{McConnell} N.~J.,  {Ma} C.-P.,  2013, \mn@doi [\apj]
  {10.1088/0004-637X/764/2/184}, \href
  {https://ui.adsabs.harvard.edu/abs/2013ApJ...764..184M} {764, 184}

\bibitem[\protect\citeauthoryear{{McLure} \& {Dunlop}}{{McLure} \&
  {Dunlop}}{2004}]{Mclure_2004}
{McLure} R.~J.,  {Dunlop} J.~S.,  2004, \mn@doi [\mnras]
  {10.1111/j.1365-2966.2004.08034.x}, \href
  {https://ui.adsabs.harvard.edu/abs/2004MNRAS.352.1390M} {352, 1390}

\bibitem[\protect\citeauthoryear{{McLure} \& {Jarvis}}{{McLure} \&
  {Jarvis}}{2002}]{Mclure_2002}
{McLure} R.~J.,  {Jarvis} M.~J.,  2002, \mn@doi [\mnras]
  {10.1046/j.1365-8711.2002.05871.x}, \href
  {https://ui.adsabs.harvard.edu/abs/2002MNRAS.337..109M} {337, 109}

\bibitem[\protect\citeauthoryear{{McMahon}, {Banerji}, {Gonzalez}, {Koposov},
  {Bejar}, {Lodieu}, {Rebolo}  \& {VHS Collaboration}}{{McMahon}
  et~al.}{2013}]{VHS}
{McMahon} R.~G.,  {Banerji} M.,  {Gonzalez} E.,  {Koposov} S.~E.,  {Bejar}
  V.~J.,  {Lodieu} N.,  {Rebolo} R.,   {VHS Collaboration} 2013, The Messenger,
  \href {https://ui.adsabs.harvard.edu/abs/2013Msngr.154...35M} {154, 35}

\bibitem[\protect\citeauthoryear{{Mej{\'\i}a-Restrepo}, {Trakhtenbrot}, {Lira},
  {Netzer}  \& {Capellupo}}{{Mej{\'\i}a-Restrepo}
  et~al.}{2016}]{Mejia-Restrepo_2016}
{Mej{\'\i}a-Restrepo} J.~E.,  {Trakhtenbrot} B.,  {Lira} P.,  {Netzer} H.,
  {Capellupo} D.~M.,  2016, \mn@doi [\mnras] {10.1093/mnras/stw568}, \href
  {https://ui.adsabs.harvard.edu/abs/2016MNRAS.460..187M} {460, 187}

\bibitem[\protect\citeauthoryear{{Nagao}, {Marconi}  \& {Maiolino}}{{Nagao}
  et~al.}{2006}]{Nagao_2006}
{Nagao} T.,  {Marconi} A.,   {Maiolino} R.,  2006, \mn@doi [\aap]
  {10.1051/0004-6361:20054024}, \href
  {https://ui.adsabs.harvard.edu/abs/2006A&A...447..157N} {447, 157}

\bibitem[\protect\citeauthoryear{{Noll}, {Kausch}, {Barden}, {Jones},
  {Szyszka}, {Kimeswenger}  \& {Vinther}}{{Noll} et~al.}{2012}]{Noll_2012}
{Noll} S.,  {Kausch} W.,  {Barden} M.,  {Jones} A.~M.,  {Szyszka} C.,
  {Kimeswenger} S.,   {Vinther} J.,  2012, \mn@doi [\aap]
  {10.1051/0004-6361/201219040}, \href
  {https://ui.adsabs.harvard.edu/abs/2012A&A...543A..92N} {543, A92}

\bibitem[\protect\citeauthoryear{{Noterdaeme} et~al.,}{{Noterdaeme}
  et~al.}{2012}]{Noterdaeme_2012}
{Noterdaeme} P.,  et~al., 2012, \mn@doi [\aap] {10.1051/0004-6361/201220259},
  \href {https://ui.adsabs.harvard.edu/abs/2012A&A...547L...1N} {547, L1}

\bibitem[\protect\citeauthoryear{{Onken} et~al.,}{{Onken}
  et~al.}{2019}]{SMSS_2019}
{Onken} C.~A.,  et~al., 2019, \mn@doi [\pasa] {10.25914/5f14eded2d116}, \href
  {https://ui.adsabs.harvard.edu/abs/2019PASA...36...33O} {36, e033}

\bibitem[\protect\citeauthoryear{{Onken}, {Bian}, {Fan}, {Wang}, {Wolf}  \&
  {Yang}}{{Onken} et~al.}{2020}]{onken_2020_J2157}
{Onken} C.~A.,  {Bian} F.,  {Fan} X.,  {Wang} F.,  {Wolf} C.,   {Yang} J.,
  2020, \mn@doi [\mnras] {10.1093/mnras/staa1635}, \href
  {https://ui.adsabs.harvard.edu/abs/2020MNRAS.496.2309O} {496, 2309}

\bibitem[\protect\citeauthoryear{{Park}, {Barth}, {Ho}  \& {Laor}}{{Park}
  et~al.}{2022}]{Park_2022}
{Park} D.,  {Barth} A.~J.,  {Ho} L.~C.,   {Laor} A.,  2022, \mn@doi [\apjs]
  {10.3847/1538-4365/ac3f3e}, \href
  {https://ui.adsabs.harvard.edu/abs/2022ApJS..258...38P} {258, 38}

\bibitem[\protect\citeauthoryear{{Pati{\~n}o {\'A}lvarez}, {Torrealba},
  {Chavushyan}, {Cruz Gonz{\'a}lez}, {Arshakian}, {Le{\'o}n Tavares}  \&
  {Popovic}}{{Pati{\~n}o {\'A}lvarez} et~al.}{2016}]{Patino_2016}
{Pati{\~n}o {\'A}lvarez} V.,  {Torrealba} J.,  {Chavushyan} V.,  {Cruz
  Gonz{\'a}lez} I.,  {Arshakian} T.,  {Le{\'o}n Tavares} J.,   {Popovic} L.,
  2016, \mn@doi [Frontiers in Astronomy and Space Sciences]
  {10.3389/fspas.2016.00019}, \href
  {https://ui.adsabs.harvard.edu/abs/2016FrASS...3...19P} {3, 19}

\bibitem[\protect\citeauthoryear{{Perrotta} et~al.,}{{Perrotta}
  et~al.}{2016}]{Perrotta_2016}
{Perrotta} S.,  et~al., 2016, \mn@doi [\mnras] {10.1093/mnras/stw1703}, \href
  {https://ui.adsabs.harvard.edu/abs/2016MNRAS.462.3285P} {462, 3285}

\bibitem[\protect\citeauthoryear{{Perrotta} et~al.,}{{Perrotta}
  et~al.}{2018}]{Perrotta_2018}
{Perrotta} S.,  et~al., 2018, \mn@doi [\mnras] {10.1093/mnras/sty2205}, \href
  {https://ui.adsabs.harvard.edu/abs/2018MNRAS.481..105P} {481, 105}

\bibitem[\protect\citeauthoryear{{Peterson} et~al.,}{{Peterson}
  et~al.}{2004}]{Peterson_2004}
{Peterson} B.~M.,  et~al., 2004, \mn@doi [\apj] {10.1086/423269}, \href
  {https://ui.adsabs.harvard.edu/abs/2004ApJ...613..682P} {613, 682}

\bibitem[\protect\citeauthoryear{{Prochaska} \& {Wolfe}}{{Prochaska} \&
  {Wolfe}}{2009}]{Prochaska_2009}
{Prochaska} J.~X.,  {Wolfe} A.~M.,  2009, \mn@doi [\apj]
  {10.1088/0004-637X/696/2/1543}, \href
  {https://ui.adsabs.harvard.edu/abs/2009ApJ...696.1543P} {696, 1543}

\bibitem[\protect\citeauthoryear{{Proga}, {Stone}  \& {Kallman}}{{Proga}
  et~al.}{2000}]{Proga_2000}
{Proga} D.,  {Stone} J.~M.,   {Kallman} T.~R.,  2000, \mn@doi [\apj]
  {10.1086/317154}, \href
  {https://ui.adsabs.harvard.edu/abs/2000ApJ...543..686P} {543, 686}

\bibitem[\protect\citeauthoryear{{Rafiee} \& {Hall}}{{Rafiee} \&
  {Hall}}{2011}]{Rafiee_2011}
{Rafiee} A.,  {Hall} P.~B.,  2011, \mn@doi [\apjs]
  {10.1088/0067-0049/194/2/42}, \href
  {https://ui.adsabs.harvard.edu/abs/2011ApJS..194...42R} {194, 42}

\bibitem[\protect\citeauthoryear{{Rakshit}, {Stalin}  \&
  {Kotilainen}}{{Rakshit} et~al.}{2020}]{Rakshit_2020}
{Rakshit} S.,  {Stalin} C.~S.,   {Kotilainen} J.,  2020, \mn@doi [\apjs]
  {10.3847/1538-4365/ab99c5}, \href
  {https://ui.adsabs.harvard.edu/abs/2020ApJS..249...17R} {249, 17}

\bibitem[\protect\citeauthoryear{{Richards}, {Vanden Berk}, {Reichard}, {Hall},
  {Schneider}, {SubbaRao}, {Thakar}  \& {York}}{{Richards}
  et~al.}{2002}]{Richards_2002}
{Richards} G.~T.,  {Vanden Berk} D.~E.,  {Reichard} T.~A.,  {Hall} P.~B.,
  {Schneider} D.~P.,  {SubbaRao} M.,  {Thakar} A.~R.,   {York} D.~G.,  2002,
  \mn@doi [\aj] {10.1086/341167}, \href
  {https://ui.adsabs.harvard.edu/abs/2002AJ....124....1R} {124, 1}

\bibitem[\protect\citeauthoryear{{Richards} et~al.,}{{Richards}
  et~al.}{2003}]{Richards_2003}
{Richards} G.~T.,  et~al., 2003, \mn@doi [\aj] {10.1086/377014}, \href
  {https://ui.adsabs.harvard.edu/abs/2003AJ....126.1131R} {126, 1131}

\bibitem[\protect\citeauthoryear{{Richards} et~al.,}{{Richards}
  et~al.}{2006}]{Richards_2006}
{Richards} G.~T.,  et~al., 2006, \mn@doi [\apjs] {10.1086/506525}, \href
  {https://ui.adsabs.harvard.edu/abs/2006ApJS..166..470R} {166, 470}

\bibitem[\protect\citeauthoryear{{Richards} et~al.,}{{Richards}
  et~al.}{2011}]{Richards_2011}
{Richards} G.~T.,  et~al., 2011, \mn@doi [\aj] {10.1088/0004-6256/141/5/167},
  \href {https://ui.adsabs.harvard.edu/abs/2011AJ....141..167R} {141, 167}

\bibitem[\protect\citeauthoryear{{Rodrigo} \& {Solano}}{{Rodrigo} \&
  {Solano}}{2020}]{SVO_Filter_Profile_Service}
{Rodrigo} C.,  {Solano} E.,  2020, in XIV.0 Scientific Meeting (virtual) of the
  Spanish Astronomical Society. p.~182

\bibitem[\protect\citeauthoryear{{Salviander}, {Shields}, {Gebhardt}  \&
  {Bonning}}{{Salviander} et~al.}{2007}]{Salviander_2007}
{Salviander} S.,  {Shields} G.~A.,  {Gebhardt} K.,   {Bonning} E.~W.,  2007,
  \mn@doi [\apj] {10.1086/513086}, \href
  {https://ui.adsabs.harvard.edu/abs/2007ApJ...662..131S} {662, 131}

\bibitem[\protect\citeauthoryear{{S{\'a}nchez-Ram{\'\i}rez}
  et~al.,}{{S{\'a}nchez-Ram{\'\i}rez} et~al.}{2016}]{Sanchez_2016}
{S{\'a}nchez-Ram{\'\i}rez} R.,  et~al., 2016, \mn@doi [\mnras]
  {10.1093/mnras/stv2732}, \href
  {https://ui.adsabs.harvard.edu/abs/2016MNRAS.456.4488S} {456, 4488}

\bibitem[\protect\citeauthoryear{{Saturni} et~al.,}{{Saturni}
  et~al.}{2018}]{Saturni_2018}
{Saturni} F.~G.,  et~al., 2018, \mn@doi [\aap] {10.1051/0004-6361/201832794},
  \href {https://ui.adsabs.harvard.edu/abs/2018A&A...617A.118S} {617, A118}

\bibitem[\protect\citeauthoryear{{Schindler} et~al.,}{{Schindler}
  et~al.}{2020}]{Schindler_2020}
{Schindler} J.-T.,  et~al., 2020, \mn@doi [\apj] {10.3847/1538-4357/abc2d7},
  \href {https://ui.adsabs.harvard.edu/abs/2020ApJ...905...51S} {905, 51}

\bibitem[\protect\citeauthoryear{{Schlafly} \& {Finkbeiner}}{{Schlafly} \&
  {Finkbeiner}}{2011}]{Schlafly_2011}
{Schlafly} E.~F.,  {Finkbeiner} D.~P.,  2011, \mn@doi [\apj]
  {10.1088/0004-637X/737/2/103}, \href
  {https://ui.adsabs.harvard.edu/abs/2011ApJ...737..103S} {737, 103}

\bibitem[\protect\citeauthoryear{{Schlafly}, {Finkbeiner}, {Schlegel},
  {Juri{\'c}}, {Ivezi{\'c}}, {Gibson}, {Knapp}  \& {Weaver}}{{Schlafly}
  et~al.}{2010}]{Schlafly_2010}
{Schlafly} E.~F.,  {Finkbeiner} D.~P.,  {Schlegel} D.~J.,  {Juri{\'c}} M.,
  {Ivezi{\'c}} {\v{Z}}.,  {Gibson} R.~R.,  {Knapp} G.~R.,   {Weaver} B.~A.,
  2010, \mn@doi [\apj] {10.1088/0004-637X/725/1/1175}, \href
  {https://ui.adsabs.harvard.edu/abs/2010ApJ...725.1175S} {725, 1175}

\bibitem[\protect\citeauthoryear{{Schlegel}, {Finkbeiner}  \&
  {Davis}}{{Schlegel} et~al.}{1998}]{Schlegel_1998}
{Schlegel} D.~J.,  {Finkbeiner} D.~P.,   {Davis} M.,  1998, \mn@doi [\apj]
  {10.1086/305772}, \href
  {https://ui.adsabs.harvard.edu/abs/1998ApJ...500..525S} {500, 525}

\bibitem[\protect\citeauthoryear{{Shen} \& {Liu}}{{Shen} \&
  {Liu}}{2012}]{Shen_2012}
{Shen} Y.,  {Liu} X.,  2012, \mn@doi [\apj] {10.1088/0004-637X/753/2/125},
  \href {https://ui.adsabs.harvard.edu/abs/2012ApJ...753..125S} {753, 125}

\bibitem[\protect\citeauthoryear{{Shen}, {Greene}, {Strauss}, {Richards}  \&
  {Schneider}}{{Shen} et~al.}{2008}]{Shen_2008}
{Shen} Y.,  {Greene} J.~E.,  {Strauss} M.~A.,  {Richards} G.~T.,   {Schneider}
  D.~P.,  2008, \mn@doi [\apj] {10.1086/587475}, \href
  {https://ui.adsabs.harvard.edu/abs/2008ApJ...680..169S} {680, 169}

\bibitem[\protect\citeauthoryear{{Shen} et~al.,}{{Shen}
  et~al.}{2011}]{Shen_2011}
{Shen} Y.,  et~al., 2011, \mn@doi [\apjs] {10.1088/0067-0049/194/2/45}, \href
  {https://ui.adsabs.harvard.edu/abs/2011ApJS..194...45S} {194, 45}

\bibitem[\protect\citeauthoryear{{Skrutskie} et~al.,}{{Skrutskie}
  et~al.}{2006}]{2MASS}
{Skrutskie} M.~F.,  et~al., 2006, \mn@doi [\aj] {10.1086/498708}, \href
  {https://ui.adsabs.harvard.edu/abs/2006AJ....131.1163S} {131, 1163}

\bibitem[\protect\citeauthoryear{{Suberlak}, {Ivezi{\'c}}  \&
  {MacLeod}}{{Suberlak} et~al.}{2021}]{Suberlak_2021}
{Suberlak} K.~L.,  {Ivezi{\'c}} {\v{Z}}.,   {MacLeod} C.,  2021, \mn@doi [\apj]
  {10.3847/1538-4357/abc698}, \href
  {https://ui.adsabs.harvard.edu/abs/2021ApJ...907...96S} {907, 96}

\bibitem[\protect\citeauthoryear{{Sulentic}, {Bachev}, {Marziani}, {Negrete}
  \& {Dultzin}}{{Sulentic} et~al.}{2007}]{Sulentic_2007}
{Sulentic} J.~W.,  {Bachev} R.,  {Marziani} P.,  {Negrete} C.~A.,   {Dultzin}
  D.,  2007, \mn@doi [\apj] {10.1086/519916}, \href
  {https://ui.adsabs.harvard.edu/abs/2007ApJ...666..757S} {666, 757}

\bibitem[\protect\citeauthoryear{{Terrazas} et~al.,}{{Terrazas}
  et~al.}{2020}]{Terrazas_2020}
{Terrazas} B.~A.,  et~al., 2020, \mn@doi [\mnras] {10.1093/mnras/staa374},
  \href {https://ui.adsabs.harvard.edu/abs/2020MNRAS.493.1888T} {493, 1888}

\bibitem[\protect\citeauthoryear{{Tsuzuki}, {Kawara}, {Yoshii}, {Oyabu},
  {Tanab{\'e}}  \& {Matsuoka}}{{Tsuzuki} et~al.}{2006}]{Tsuzuki_2006}
{Tsuzuki} Y.,  {Kawara} K.,  {Yoshii} Y.,  {Oyabu} S.,  {Tanab{\'e}} T.,
  {Matsuoka} Y.,  2006, \mn@doi [\apj] {10.1086/506376}, \href
  {https://ui.adsabs.harvard.edu/abs/2006ApJ...650...57T} {650, 57}

\bibitem[\protect\citeauthoryear{{Vanden Berk} et~al.,}{{Vanden Berk}
  et~al.}{2001}]{VandenBerk_2001}
{Vanden Berk} D.~E.,  et~al., 2001, \mn@doi [\aj] {10.1086/321167}, \href
  {https://ui.adsabs.harvard.edu/abs/2001AJ....122..549V} {122, 549}

\bibitem[\protect\citeauthoryear{{Vanden Berk} et~al.,}{{Vanden Berk}
  et~al.}{2004}]{VandenBerk_2004}
{Vanden Berk} D.~E.,  et~al., 2004, \mn@doi [\apj] {10.1086/380563}, \href
  {https://ui.adsabs.harvard.edu/abs/2004ApJ...601..692V} {601, 692}

\bibitem[\protect\citeauthoryear{{Vernet} et~al.,}{{Vernet}
  et~al.}{2011}]{Vernet_2011_Xshooter}
{Vernet} J.,  et~al., 2011, \mn@doi [\aap] {10.1051/0004-6361/201117752}, \href
  {https://ui.adsabs.harvard.edu/abs/2011A&A...536A.105V} {536, A105}

\bibitem[\protect\citeauthoryear{{Vestergaard}}{{Vestergaard}}{2002}]{Vestergaard_2002}
{Vestergaard} M.,  2002, \mn@doi [\apj] {10.1086/340045}, \href
  {https://ui.adsabs.harvard.edu/abs/2002ApJ...571..733V} {571, 733}

\bibitem[\protect\citeauthoryear{{Vestergaard} \& {Osmer}}{{Vestergaard} \&
  {Osmer}}{2009}]{Vestergaard_2009}
{Vestergaard} M.,  {Osmer} P.~S.,  2009, \mn@doi [\apj]
  {10.1088/0004-637X/699/1/800}, \href
  {https://ui.adsabs.harvard.edu/abs/2009ApJ...699..800V} {699, 800}

\bibitem[\protect\citeauthoryear{{Vestergaard} \& {Peterson}}{{Vestergaard} \&
  {Peterson}}{2006}]{Vestergaard_2006}
{Vestergaard} M.,  {Peterson} B.~M.,  2006, \mn@doi [\apj] {10.1086/500572},
  \href {https://ui.adsabs.harvard.edu/abs/2006ApJ...641..689V} {641, 689}

\bibitem[\protect\citeauthoryear{{Vestergaard} \& {Wilkes}}{{Vestergaard} \&
  {Wilkes}}{2001}]{Vestergaard_2001}
{Vestergaard} M.,  {Wilkes} B.~J.,  2001, \mn@doi [\apjs] {10.1086/320357},
  \href {https://ui.adsabs.harvard.edu/abs/2001ApJS..134....1V} {134, 1}

\bibitem[\protect\citeauthoryear{{Virtanen} et~al.,}{{Virtanen}
  et~al.}{2020}]{Scipy_2020}
{Virtanen} P.,  et~al., 2020, \mn@doi [Nature Methods]
  {10.1038/s41592-019-0686-2}, \href
  {https://ui.adsabs.harvard.edu/abs/2020NatMe..17..261V} {17, 261}

\bibitem[\protect\citeauthoryear{{Wang} et~al.,}{{Wang}
  et~al.}{2009}]{Wang_2009}
{Wang} J.-G.,  et~al., 2009, \mn@doi [\apj] {10.1088/0004-637X/707/2/1334},
  \href {https://ui.adsabs.harvard.edu/abs/2009ApJ...707.1334W} {707, 1334}

\bibitem[\protect\citeauthoryear{{Wang}, {Zhou}, {Yuan}  \& {Wang}}{{Wang}
  et~al.}{2012}]{Wang_2012}
{Wang} H.,  {Zhou} H.,  {Yuan} W.,   {Wang} T.,  2012, \mn@doi [\apjl]
  {10.1088/2041-8205/751/2/L23}, \href
  {https://ui.adsabs.harvard.edu/abs/2012ApJ...751L..23W} {751, L23}

\bibitem[\protect\citeauthoryear{{Wang} et~al.,}{{Wang}
  et~al.}{2021}]{Wang_2021_z7.642}
{Wang} F.,  et~al., 2021, \mn@doi [\apjl] {10.3847/2041-8213/abd8c6}, \href
  {https://ui.adsabs.harvard.edu/abs/2021ApJ...907L...1W} {907, L1}

\bibitem[\protect\citeauthoryear{{Wang} et~al.,}{{Wang}
  et~al.}{2022}]{Wang_2021}
{Wang} S.,  et~al., 2022, \mn@doi [\apj] {10.3847/1538-4357/ac3a69}, \href
  {https://ui.adsabs.harvard.edu/abs/2022ApJ...925..121W} {925, 121}

\bibitem[\protect\citeauthoryear{Waskom}{Waskom}{2021}]{Waskom2021_seaborn}
Waskom M.~L.,  2021, \mn@doi [Journal of Open Source Software]
  {10.21105/joss.03021}, 6, 3021

\bibitem[\protect\citeauthoryear{{Wolfe}, {Gawiser}  \& {Prochaska}}{{Wolfe}
  et~al.}{2005}]{Wolfe_2005}
{Wolfe} A.~M.,  {Gawiser} E.,   {Prochaska} J.~X.,  2005, \mn@doi [\araa]
  {10.1146/annurev.astro.42.053102.133950}, \href
  {https://ui.adsabs.harvard.edu/abs/2005ARA&A..43..861W} {43, 861}

\bibitem[\protect\citeauthoryear{{Woo}, {Le}, {Karouzos}, {Park}, {Park},
  {Malkan}, {Treu}  \& {Bennert}}{{Woo} et~al.}{2018}]{Woo_2018}
{Woo} J.-H.,  {Le} H. A.~N.,  {Karouzos} M.,  {Park} D.,  {Park} D.,  {Malkan}
  M.~A.,  {Treu} T.,   {Bennert} V.~N.,  2018, \mn@doi [\apj]
  {10.3847/1538-4357/aabf3e}, \href
  {https://ui.adsabs.harvard.edu/abs/2018ApJ...859..138W} {859, 138}

\bibitem[\protect\citeauthoryear{{Xu}, {Bian}, {Shen}, {Zuo}, {Fan}  \&
  {Zhu}}{{Xu} et~al.}{2018}]{Xu_2018}
{Xu} F.,  {Bian} F.,  {Shen} Y.,  {Zuo} W.,  {Fan} X.,   {Zhu} Z.,  2018,
  \mn@doi [\mnras] {10.1093/mnras/sty1763}, \href
  {https://ui.adsabs.harvard.edu/abs/2018MNRAS.480..345X} {480, 345}

\bibitem[\protect\citeauthoryear{{Yao} et~al.,}{{Yao} et~al.}{2019}]{Yao_2019}
{Yao} S.,  et~al., 2019, \mn@doi [\apjs] {10.3847/1538-4365/aaef88}, \href
  {https://ui.adsabs.harvard.edu/abs/2019ApJS..240....6Y} {240, 6}

\bibitem[\protect\citeauthoryear{{Y{\`e}che}, {Palanque-Delabrouille}, {Baur}
  \& {du Mas des Bourboux}}{{Y{\`e}che} et~al.}{2017}]{Yeche_2017}
{Y{\`e}che} C.,  {Palanque-Delabrouille} N.,  {Baur} J.,   {du Mas des
  Bourboux} H.,  2017, \mn@doi [\jcap] {10.1088/1475-7516/2017/06/047}, \href
  {https://ui.adsabs.harvard.edu/abs/2017JCAP...06..047Y} {2017, 047}

\bibitem[\protect\citeauthoryear{{Yong}, {Webster}, {King}, {Bate}, {Labrie}
  \& {O'Dowd}}{{Yong} et~al.}{2020}]{Yong_2020}
{Yong} S.~Y.,  {Webster} R.~L.,  {King} A.~L.,  {Bate} N.~F.,  {Labrie} K.,
  {O'Dowd} M.~J.,  2020, \mn@doi [\mnras] {10.1093/mnras/stz3074}, \href
  {https://ui.adsabs.harvard.edu/abs/2020MNRAS.491.1320Y} {491, 1320}

\bibitem[\protect\citeauthoryear{{York} et~al.,}{{York}
  et~al.}{2000}]{York_2000_SDSS}
{York} D.~G.,  et~al., 2000, \mn@doi [\aj] {10.1086/301513}, \href
  {https://ui.adsabs.harvard.edu/abs/2000AJ....120.1579Y} {120, 1579}

\bibitem[\protect\citeauthoryear{{van der Walt}, {Colbert}  \&
  {Varoquaux}}{{van der Walt} et~al.}{2011}]{Numpy_2011}
{van der Walt} S.,  {Colbert} S.~C.,   {Varoquaux} G.,  2011, \mn@doi
  [Computing in Science and Engineering] {10.1109/MCSE.2011.37}, \href
  {https://ui.adsabs.harvard.edu/abs/2011CSE....13b..22V} {13, 22}

\makeatother
\end{thebibliography}

%\begin{thebibliography}{99}
%\end{thebibliography}

%%%%%%%%%%%%%%%%%%%%%%%%%%%%%%%%%%%%%%%%%%%%%%%%%%

%%%%%%%%%%%%%%%%% APPENDICES %%%%%%%%%%%%%%%%%%%%%

\appendix

\section{Significance of the Balmer Continuum} \label{sec:appendix-Balmer}
In Section \ref{sec:cont-fitting}, we discussed our pseudo-continuum model which includes two components: a power-law and \feii\ flux. The Balmer continuum is also often considered as part of a QSO pseudo-continuum model. However, we found that for the sources in our XQ-100 sample, the broad emission-line models are not strongly affected by either the inclusion or exclusion of the Balmer continuum. We refer to similar studies of QSO spectra and use the following model for the Balmer continuum \citep[e.g.,][]{Grandi_1982, Dietrich_2002, Wang_2009, Kovacevic_2014}:
\begin{equation}
\begin{aligned}
        F_{\rm{Bal}}(\lambda; F_{\rm{0}}, T_{\rm{e}}, \tau_{\lambda}) = F_{\rm{Bal,0}} \, B_{\lambda}(\lambda, T_{\rm{e}})(1-e^{-\tau_{\lambda}}); && \lambda \leq \lambda_{\rm{BE}}\,,
\end{aligned}
\end{equation}
where $F_{\rm{Bal, 0}}$ is the normalisation, $B_{\lambda}(\lambda, T_{\rm{e}})$ is the Planck blackbody with uniform electron temperature $T_{\rm{e}}$, and $\tau_{\lambda}$ is the optical depth. This Balmer continuum is defined at wavelengths shorter than the Balmer edge at rest-frame $\lambda_{\rm{BE}} \equiv 3646$ \AA. Here, we quantify the difference in our results by including the Balmer continuum in Table \ref{tab:Balmer_difference}, which measures the effect in the resulting \mgii\ model properties. From the entire sample of 100 QSOs, we quantify the effect in terms of the mean residual and its standard deviation. The change in each line property resulting from pseudocontinuum models with the Balmer continuum is typically no larger than 1\%. With reference to the mean uncertainties of the measured emission-line properties in our sample, the inclusion of the Balmer continuum has a statistically insignificant effect on all measured \mgii\ line properties and there is no observable systematic bias to emission-line models by removing the Balmer continuum from our pseudocontinuum model. However, the absence of the Balmer model can affect the slope of the power-law continuum and amplitude of the \feii\ models.

\begingroup
\begin{table}
\centering{
\caption {\label{tab:Balmer_difference} Mean difference and standard deviation in the measured line properties of the \mgii\ line between including and excluding the Balmer continuum flux contribution. Refer to Table \ref{tab:XQ100_line_properties} for a description of each line property. The monochromatic luminosity at 3000\AA, L$_{\rm{3000}}$, has units of erg s$^{-1}$ and is represented on a log-scale as with the integrated line luminosity, iLuminosity. For comparison, we show the mean value measured in the XQ-100 sample for each line property and the mean uncertainty.}  
\begin{tabular}{lcc}
\hline \hline
\mgii\ Property & Balmer Residual & Sample Mean\\
\hline
FWHM & \hphantom{~~~}$54\pm189$ & \hphantom{~}$3633\pm237$\\
Sigma & \hphantom{~~~}$16\pm146$ & \hphantom{~}$2821\pm286$\\
Blueshift & \hphantom{~~~}$7\pm45$ & $-240\pm99$\hphantom{~}\\
EW & \hphantom{~}$0.2\pm1.0$ & \hphantom{~~}$30\pm2$\\
pWavelength & $-0.10\pm0.58$\hphantom{~} & $2803\pm1$\hphantom{~~}\\
iLuminosity & \hphantom{~}$0.00\pm0.01$ & \hphantom{~~}$44.78\pm0.04$ \\
L$_{\rm{3000}}$ & \hphantom{~}$0.00\pm0.01$ & \hphantom{~~}$46.78\pm0.01$\\
\hline \hline
\end{tabular}
}
\end{table}
\endgroup
%\hphantom{~~~~~~}

\section{Black hole mass comparisons} \label{sec:appendix-bhcompare}
In this section, we discuss outliers between black hole masses determined from different emission-lines and variations in mass measurements between different assumptions for the \feii\ model. Among those with a greater than $0.3$ dex difference between the \mgii-based single-epoch virial mass estimate and the averaged mass from all measured emission-lines as flagged by the ``Mbh\_Flag'', we focus on two targets with the highest mass discrepancy that are not flagged by any quality flags, PSS J0121$+$0347 and BR J0714--6455, with $\log\left(\rm{M_{mean}}/\rm{M_{\mgii}}\right)$ = $-0.36$ and $0.31$, respectively. As another example, we show SDSSJ1202--0054, which was identified as an outlier in Section \ref{sec:sample-props} due to greater than 0.5 dex difference between the \hbeta\ and \civ-based black hole masses.

\begin{figure*}
\begin{tabular}{c}
  \includegraphics[width=\textwidth]{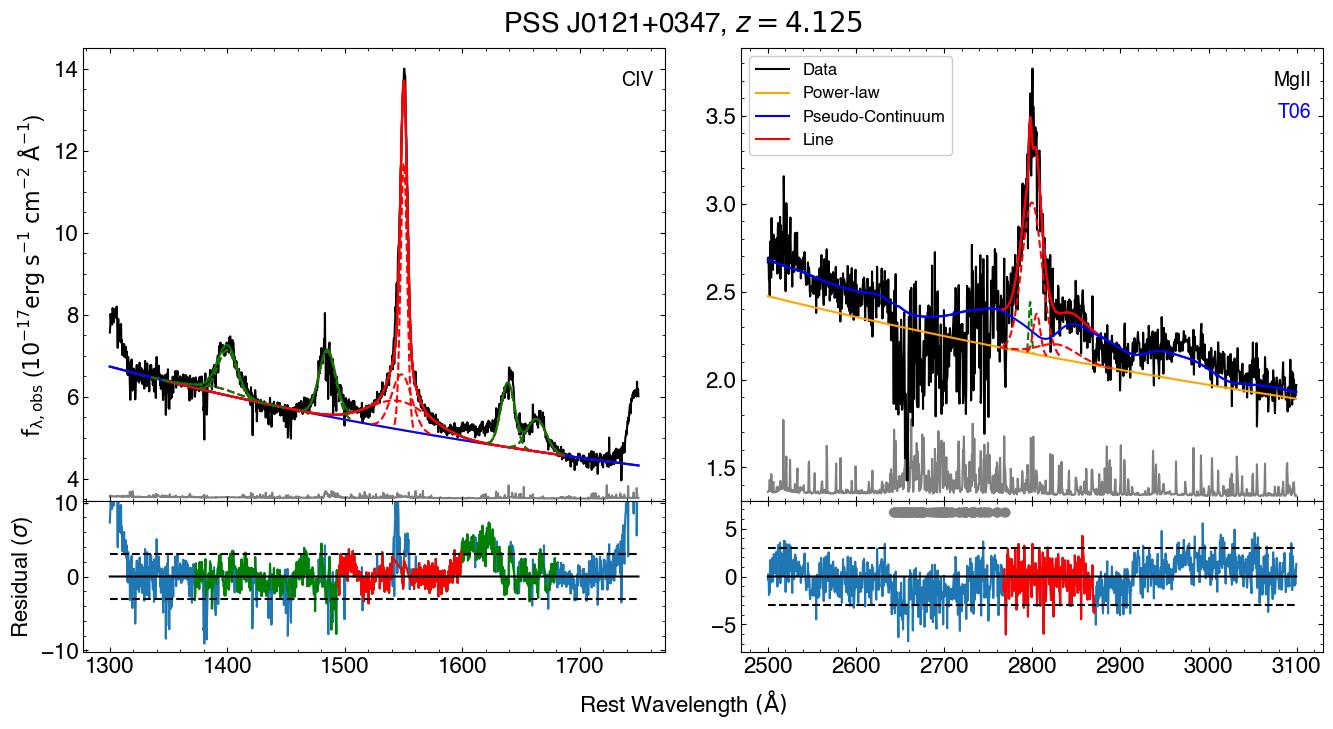} \\   
  \includegraphics[width=\textwidth]{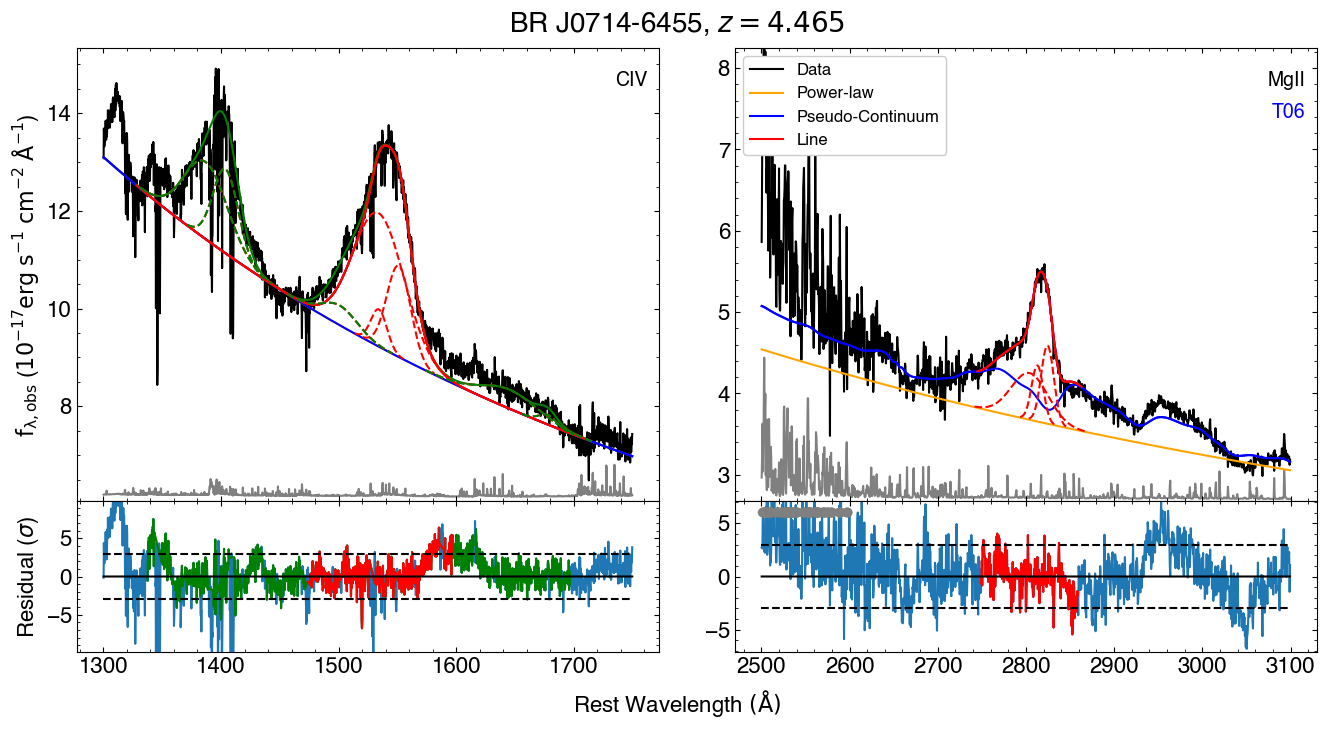} 
\end{tabular}
\caption{Example models of the \civ\ and \mgii\ emission-lines from PSS J0121$+$0347 and BR J0714--6455, outliers identified by the ``Mbh\_Flag'' which indicates when the \mgii-based virial mass estimate differs from the mean mass estimated from all measured emission-lines by over 0.3 dex. Refer to Figure \ref{fig:Example_fits} for a description of the plotted elements.} 
\label{fig:outlier-plots}
\end{figure*}

In Figure \ref{fig:outlier-plots}, we present example emission-line models of \civ\ and \mgii\ for the ``Mbh\_Flag'' outliers PSS J0121$+$0347 and BR J0714--6455. In the case of PSS J0121$+$0347, the \civ\ line profile is sharply peaked. As seen from the residual on both sides of the peak, the intrinsic line profile is possibly broader than the model suggests, but complexities of the line near its peak could not be adequately modeled by three Gaussian components. The narrowest \civ\ component has a FWHM of $\sim$540 km s$^{-1}$. Although some studies argue in favour of narrow-line subtraction for \civ\ \citep{baskin_laor_2005}, we do not make this consideration in this study.

For BR J0714--6455, the \civ\ line profile may be broadened by non-virial components, as evidenced by the strong asymmetry present in both \civ\ and the \siiv$+$\oiv\ complex. The relative velocity shift between the two lines is $\Delta v(\rm{\civ-\mgii}) \sim 3300$ km s$^{-1}$, which is in the top 10\% highest relative velocity shifts measured in our sample. The mean mass discrepancy and its standard deviation for targets with $\Delta v(\rm{\civ-\mgii}) > 3000$ km s$^{-1}$ is $\log\left(\rm{M_{\civ}}/\rm{M_{\mgii}}\right) = 0.25 \pm 0.14$. At lower velocity shifts, this discrepancy is $\log\left(\rm{M_{\civ}}/\rm{M_{\mgii}}\right) = 0.01 \pm 0.14$, indicating the potential for outflows to preferentially bias the \civ\ line profile and resulting mass measurements relative to other lines.

\begin{figure*}
  \includegraphics[width=\textwidth]{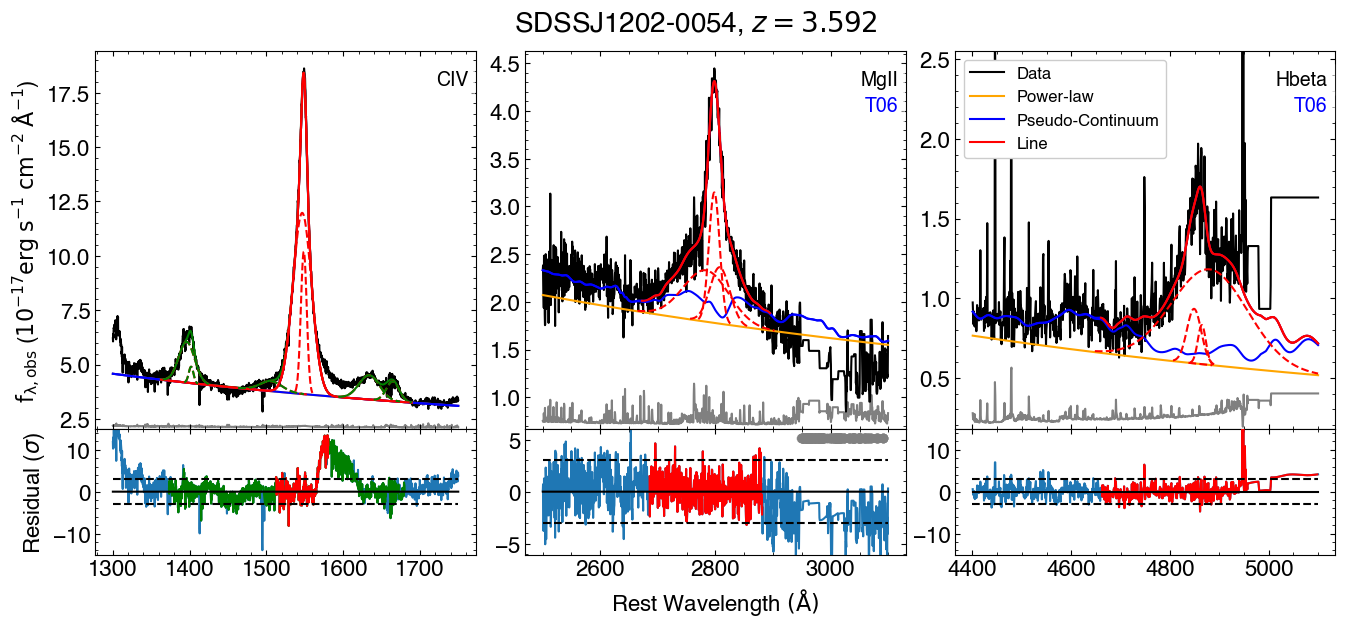}
\caption{Example models of the \civ, \mgii, and \hbeta\ emission-lines from SDSSJ1202--0054, an outlier identified by a discrepancy in the \civ\ and \hbeta-based virial mass estimates, $\log\left(\rm{M_{\civ}}/\rm{M_{\hbeta}}\right) \approx -0.8$. Refer to Figure \ref{fig:Example_fits} for a description of the plotted elements.} \label{fig:Hbeta_outlier}
\end{figure*}

In Figure \ref{fig:Hbeta_outlier}, we present example emission-line models of all three broad lines for SDSSJ1202--0054, one of the outliers identified due to the discrepancy between \civ\ and \hbeta-based virial mass estimates, $\log\left(\rm{M_{\civ}}/\rm{M_{\hbeta}}\right) \approx -0.8$. Similar to PSS J0121$+$0347, the \civ\ line model is sharply peaked with FWHM $\sim$1750 km s$^{-1}$ for the narrowest \civ\ component. However, the \hbeta\ line profile is truncated and its broad component is poorly constrained. For this target, the \mgii\ virial mass estimate is intermediate between the \hbeta\ and \civ-based estimates.

From Section \ref{sec:spectral_fitting}, we discussed how properties of the \mgii\ and \hbeta\ emission-line models are sensitive to the choice of \feii\ template. In this study, our reported line properties are averaged over measurements based on applying spectral decomposition with four different UV and optical \feii\ templates. However, we note that one should always be careful about systematic effects introduced by the choice of templates, particularly if the template is different from the one used in the virial mass calibration \citep[e.g.,][]{Woo_2018, Schindler_2020}. For instance, \mgii\ line models obtained using the VW01 template to model the underlying \feii\ emission has been shown to be biased towards broader line profiles, resulting in higher black hole mass estimates \citep[e.g.,][]{Schindler_2020}.

\begin{figure*}
  \includegraphics[width=\textwidth]{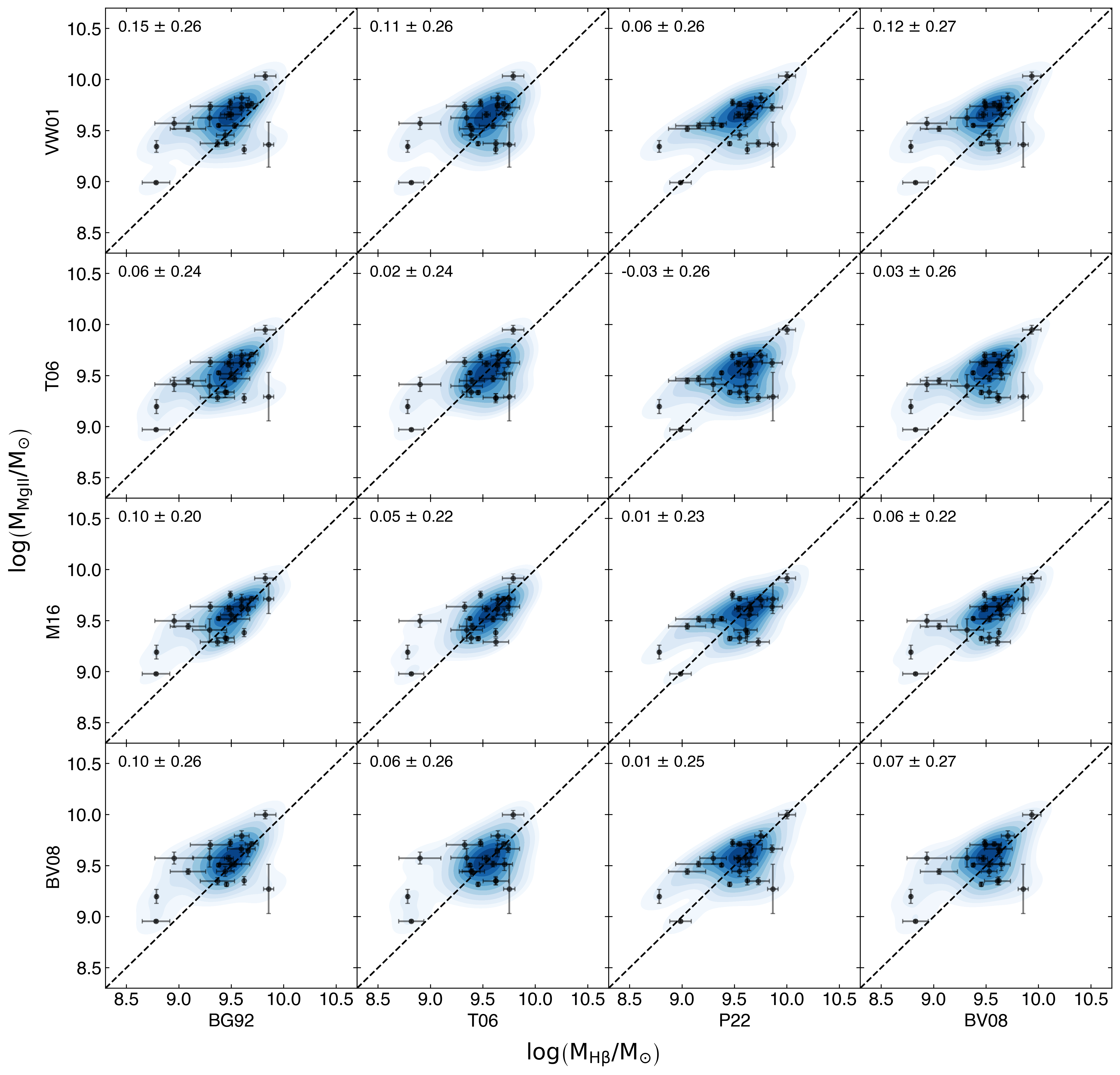}
\caption{Black hole mass comparison, $\log\left(\rm{M_{BH}}/\rm{M_{\odot}}\right)$, between different \feii\ templates models for both \mgii\ and \hbeta. The top left value displays the mean residual, $\log\left(\rm{M_{\mgii}}/\rm{M_{\hbeta}}\right)$, and standard deviation between mass measurements. The \feii\ templates are labelled along the axes. The blue shaded contours represent the two-dimensional continuous probability density distribution calculated with a kernel density estimator \citep{Waskom2021_seaborn}. Each subsequent contour level marks density iso-proportions increasing by an additional 10\% up to 90\% enclosed.} \label{fig:Mbh_Fe_Compare}
\end{figure*}

In Figure \ref{fig:Mbh_Fe_Compare}, we compare black hole mass, $\log{(\rm{M_{BH}}/\rm{M_\odot})}$, measurement pairs between \mgii\ and \hbeta-based single-epoch virial estimates. We split the panels by the \feii\ template, resulting in a $4\times4$ grid of panels for the four UV templates and four optical templates, described in Section \ref{sec:cont-fitting}, used to constrain the pseudo-continuum model in the vicinity of each emission-line. We show the mean residual, $\log\left(\rm{M_{\mgii}}/\rm{M_{\hbeta}}\right)$, and its standard deviation on the top left corner of each panel. The majority of single-epoch virial mass calibrations are based on \mgii\ and \hbeta\ emission-line models obtained using the original VW01 and BG92 templates \citep[e.g.,][]{Mclure_2004, Vestergaard_2006}, so combination of these two templates may be expected to present the tightest relationship. However, it is not apparent in this comparison that those measurements are more consistent than the results from any other pair of templates. Thus in our study, we report the line properties from averaging measurements of resulting line models based on applying spectral decomposition with each \feii\ template.

\begin{figure*}
  \includegraphics[width=\textwidth]{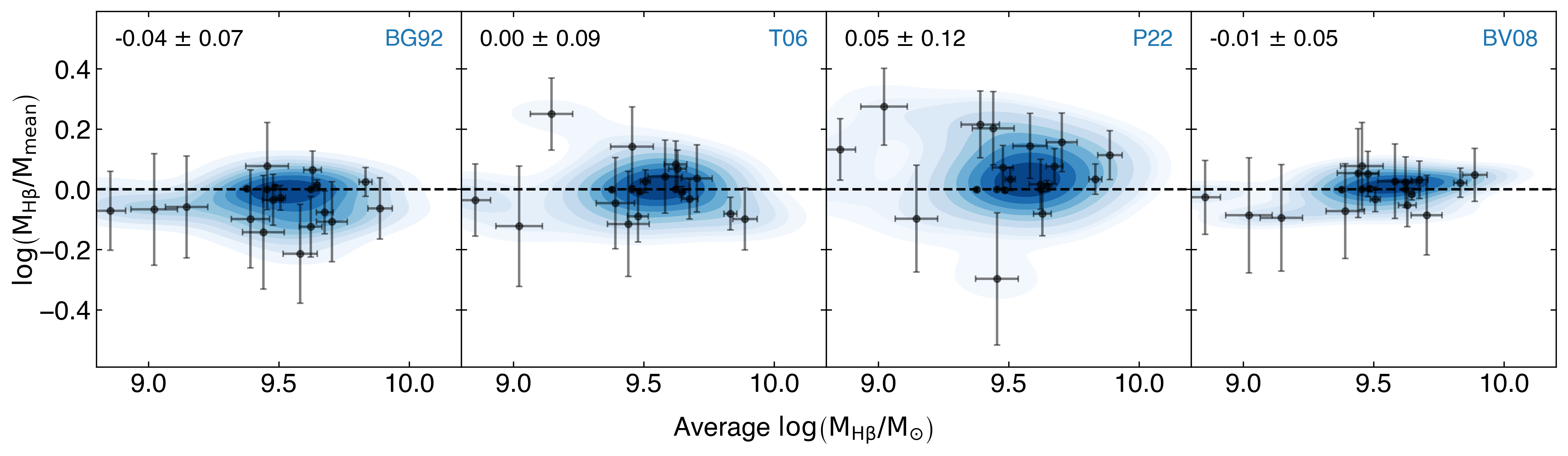}
\caption{Black hole mass comparison, $\log\left(\rm{M_{\hbeta}}/\rm{M_{mean}}\right)$, between different \feii\ templates models for \hbeta, compared to the average \hbeta-based mass estimate. The top left value displays the mean residual and its standard deviation between mass measurements. The \feii\ template is identified in the top-right corner of each panel. The blue shaded contours represent the two-dimensional continuous probability density distribution calculated with a kernel density estimator \citep{Waskom2021_seaborn}. Each subsequent contour level marks density iso-proportions increasing by an additional 10\% up to 90\% enclosed.} \label{fig:Mbh_Hbeta_Compare}
\end{figure*}

\begin{figure*}
  \includegraphics[width=\textwidth]{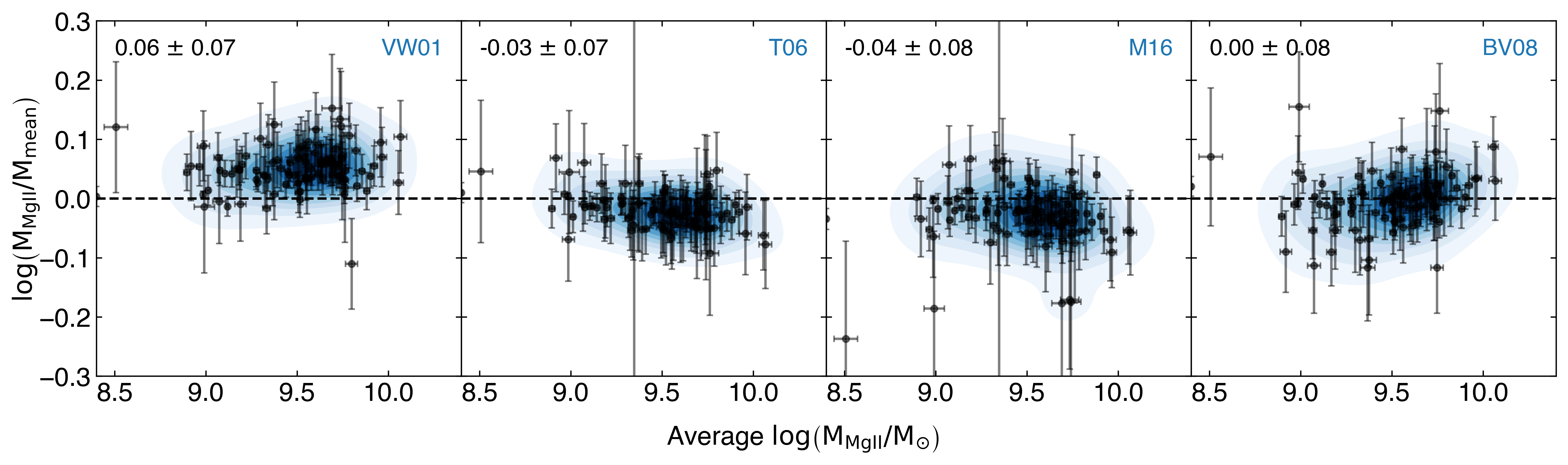}
\caption{Black hole mass comparison, $\log\left(\rm{M_{\mgii}}/\rm{M_{mean}}\right)$, between different \feii\ templates models for \mgii, compared to the average \mgii-based mass estimate. The top left value displays the mean residual and its standard deviation between mass measurements. The \feii\ template is identified in the top-right corner of each panel. The blue shaded contours represent the two-dimensional continuous probability density distribution calculated with a kernel density estimator \citep{Waskom2021_seaborn}. Each subsequent contour level marks density iso-proportions increasing by an additional 10\% up to 90\% enclosed.} \label{fig:Mbh_MgII_Compare}
\end{figure*}

In order to examine the black hole mass estimate deviation caused by each individual \feii\ template from the measured average, we present Figures \ref{fig:Mbh_Hbeta_Compare} and \ref{fig:Mbh_MgII_Compare}, which compare measurements of \hbeta\ and \mgii\ against their respective mean black hole mass estimates. In each panel, the residual, $\log\left(\rm{M_{line}}/\rm{M_{mean}}\right)$, is plotted with its mean shown on the top left along with its standard deviation. From Figure \ref{fig:Mbh_Hbeta_Compare}, we observe that all templates of the underlying \feii\ emission result in statistically similar black hole mass estimates, albeit with a fairly large scatter in this high-redshift sample due to the lower SNR of the spectrum around \hbeta\ and often truncated line profile. In Figure \ref{fig:Mbh_MgII_Compare}, we observe the effect seen in \citet{Schindler_2020} that the VW01 template tends to result in higher FWHM and overestimated black hole masses compared to other templates. However, at approximately $1\sigma$, we note that the effect is fairly insignificant with respect to the scatter, likely because our VW01 template is a modified version of the original template from \citet{Vestergaard_2001}, as elaborated upon in Section \ref{sec:cont-fitting}. With these comparisons, it is reasonable to assume that, at least for black hole mass estimates, all of the \feii\ templates tested in this study have similar effects on the resulting line properties and there are minimal observable systematics between them.

%\section{Additional Tables and Figures}

%%%%%%%%%%%%%%%%%%%%%%%%%%%%%%%%%%%%%%%%%%%%%%%%%%

% Don't change these lines
\bsp	% typesetting comment
\label{lastpage}
\end{document}